\documentclass{aa}  

\usepackage{hyperref}
\usepackage{graphicx}
\usepackage{txfonts}
\usepackage{amsmath}
\usepackage{gensymb}
\usepackage{xcolor}
\usepackage{subfig}
\usepackage{caption}
\usepackage{array}
\usepackage{siunitx}

\makeatletter
\renewcommand{\fnum@table}{\textbf{Table~\thetable}}
\makeatother

\makeatletter
\renewcommand{\fnum@figure}{\textbf{Fig.~\thefigure}}
\makeatother

\usepackage{textcomp}
\begin{document}

   \title{Unbiased estimates of the shapes of haloes using the positions of satellite galaxies}

  \author{A. Herle \inst{1}
  \and N. E. Chisari \inst{2, 1}
  \and H. Hoekstra \inst{1} 
  \and R. J. McGibbon \inst{1}
  \and J. Schaye \inst{1} 
  \and M. Schaller \inst{3, 1}
  \and R. Kugel \inst{1}  }

   \institute{Leiden Observatory, Leiden University, Niels Bohrweg 2, 2333 CA, Leiden, the Netherlands\\
              \email{herle@strw.leidenuniv.nl}
         \and
              Institute for Theoretical Physics, Utrecht University, Princetonplein 5, 3584 CC, Utrecht, the Netherlands
            \and 
                Lorentz Institute for Theoretical Physics, Leiden University, PO box 9506, 2300 RA Leiden, the Netherlands\\
             }


 
  \abstract
   {The shapes of dark matter haloes are sensitive to both cosmology and baryon physics, but are difficult to measure observationally. A promising way to constrain them is to use the positions of satellite galaxies as tracers of the underlying dark matter, but there are typically too few galaxies per halo for reliable shape estimates, resulting in biased shapes. We present a method to model sampling noise to correct for the shape bias. We compare our predicted median shape bias with that obtained from the FLAMINGO suite of simulations and find reasonable agreement. We check that our results are robust to resolution effects and baryonic feedback. We also explore the validity of our bias correction at various redshifts and we discuss how our method might be applied to observations in the future. We show that median projected halo axis ratios are on average biased low by 0.31 when they are traced by only 5 satellites. Using the satellite galaxies, the projected host halo axis ratio can be corrected with a residual bias of $\sim$ 0.1, by accounting for sampling bias. Hence, about two-thirds of the projected axis ratio bias can be explained by sampling noise. This enables the statistical measurement of halo shapes at lower masses than previously possible. Our method will also allow improved estimates of halo shapes in cosmological simulations using fewer particles than currently required.}


   \maketitle
%

\section{Introduction}
In the current cosmological paradigm, structures form hierarchically through accretion from their environment or through mergers of smaller collapsed objects. As most of the matter is believed to be cold dark matter, this process is dominated by gravity, and thus can be captured well in large cosmological simulations. The growth of cosmic structures is closely connected to the formation and evolution of galaxies, but the observable properties of the latter are more difficult to predict from first principles. Nonetheless, linking the observable properties of galaxies to their surrounding dark matter haloes is a key objective of galaxy formation models.

An important prediction from hierarchical structure formation is that the resulting collapsed regions, or dark matter haloes, are triaxial, and appear elliptical in projection \citep{Dubinski1991, Jing2002}. Since haloes grow from mergers, their shapes are sensitive to halo assembly over cosmic timescales \citep[e.g][]{Sereno2018}. For instance, haloes that formed earlier tend to be more spherical at the current epoch than those that formed late \citep{Ragone-Figueroa2010, Jeeson-Daniel2011, Lau2021}. As a result, the distribution of halo shapes depends on the cosmological model and thus can be used to constrain cosmological parameters \citep{Ho2006}. However, the nature of dark matter can also influence the shapes of haloes, as self-interacting dark matter produces haloes that are rounder towards the centre than cold dark matter \citep[e.g.][]{Peter2013, Robertson2019, Gonzalez2024}, while baryonic feedback processes can have a similar impact \citep[e.g.][and references therein]{Bryan2013}.

Correlations between the orientations of haloes are also expected, because material is accreted through filaments, and large-scale tidal interactions align them \citep[e.g.][]{Catelan2001, Hirata2004, Hirata2007}. This has implications for the correlations between the observed shapes of galaxies: these intrinsic alignments (IA) are an important contaminant for cosmological weak lensing surveys (see \citet{Joachimi2015} for a review). Studies have also shown that the shapes of the halo and central galaxy are connected \citep{Okumura2009, Schneider2010, Fortuna2021}. Hydrodynamical simulations can be used to explore this connection, but predicting the level of alignment of the star light in hydrodynamical simulations remains challenging \citep[e.g.][]{Velliscig2015a, Chisari2017}. It requires both high resolution and a large simulation volume, while observational constraints are limited \citep[e.g.][]{vanUitert2017a, Zhou2023, Shi2024}.

Constraining the distribution of halo shapes observationally, and linking these to robust predictions from cosmological N-body simulations is thus of great interest. A key challenge, however, is that it is difficult to study the dark matter haloes themselves. One way is to use weak gravitational lensing, which can provide direct estimates for individual massive clusters \citep{Hoekstra1998, Oguri2010}. The study of lower mass systems, however, requires stacking the signals of large samples of lenses, which complicates the interpretation \citep{Hoekstra2004, Mandelbaum2006, vanUitert2017b, Georgiou2021, Schrabback2021}.

Another prediction of the cold dark matter paradigm is the presence of substructures within haloes. If these satellite overdensities are massive enough, they can contain galaxies. An example are the galaxies in a galaxy cluster or group. The positions of these satellite galaxies may contain information about the shape of the host halo. Since dark matter haloes align much stronger than galaxies, linking the galaxies that we observe with their host dark matter is very useful, especially in the context of IA. For instance, \citet{Agustsson2006} showed that the satellites preferentially lie along the major axis of the ellipse traced by the halo in projection, and \citep{Shao2016} showed that satellites can trace the host's dark matter well, especially in the outer regions of the halo. This opens up the possibility of using the satellites to trace the underlying dark matter distribution for individual (high mass) objects \citep{Zhou2023, vanUitert2017a}. However, whether these satellites are an unbiased tracer of the dark matter halo shape requires further study, which is the aim of this paper. 

The most massive clusters may contain over a thousand satellite galaxies, but the majority of these are very dim and cannot be seen unless very deep observations are available. This implies that in practice, halo shapes need to be inferred using small samples of satellite positions. As a consequence, the estimate of the shape is expected to be biased due to sampling noise, an effect that has been studied in the context of shape measurements of sources for weak lensing \citep{Melchior2012}. Here, we use results from FLAMINGO \citep[Full-hydro Large-scale structure simulations with All-sky Mapping for the Interpretation of Next Generation Observations;][]{Schaye2023, Kugel2023}, a state-of-the-art suite of cosmological hydrodynamical simulations, to study the prospects of using the positions of satellite galaxies as a probe for the shape of the host halo. Specifically, we focus on the challenges posed by sampling bias.

This paper is structured as follows. In Section \ref{sect:flamingo}, we detail the cosmological simulations used in this work. Section \ref{sect:methods} details our approach to estimating the bias from sampling noise on the shapes of haloes, while we compare with the simulations in Section \ref{sect:results}. In Section \ref{sect:misalignment} we study the radial and mass dependence of the halo orientation angle to link the galaxies to their host dark matter. We discuss our results in Section \ref{sect:disc_conc}.

\section{FLAMINGO Simulations}
\label{sect:flamingo}

Throughout this paper, we used the output of FLAMINGO, which is a Virgo consortium project presented in detail in \citet{Schaye2023}. FLAMINGO is a large suite of cosmological structure formation simulations that encompass variations in cosmology, baryonic feedback and numerical resolution. These simulations are particularly useful for our application because the large box sizes result in a very large number of well-resolved haloes at the high mass end. Moreover, the different runs allow us to study variations due to resolution, redshift and feedback model. Details of the runs used in this work are shown in Table \ref{table:sim_details}. A novel feature of FLAMINGO is the use of machine learning for the calibration of the sub-grid prescriptions of stellar and AGN feedback to the observed low redshift cluster gas fractions and galaxy stellar mass function, described in detail by \citet{Kugel2023}. 

The simulations were run using the SWIFT hydrodynamics code \citep{Schaller2024} with the SPHENIX \citep{Borrow2022} implementation for smoothed particle hydrodynamics, specifically designed for galaxy formation simulations. Neutrinos are modelled using the method proposed by \citet{Elbers2021} that allows a consistent treatment of both small and large neutrino masses.

Radiative cooling and heating are implemented element-by-element as described by \citet{Ploeckinger2020}, with a new treatment of rapidly cooling gas. Gas particles are stochastically converted into star particles with a pressure-dependent star formation rate as described by \citet{Schaye2008}. Time-dependent stellar mass-loss due to stellar winds and supernovae are implemented as in \citet{Wiersma2009} \citep[with modifications described in][]{Schaye2015}. Stellar energy feedback is implemented by kicking pairs of young star particles in random but opposite directions \citep{DallaVecchia2008, Chaikin2023}. Gas accretion onto supermassive black holes and the resulting thermal AGN feedback is implemented as prescribed by \citet{Booth2009}, whereas the kinetic jet feedback used in the Jets simulation is implemented as described by \citet{Husko2022}.

\begin{figure*}
    \centering
    \includegraphics[scale=0.5]{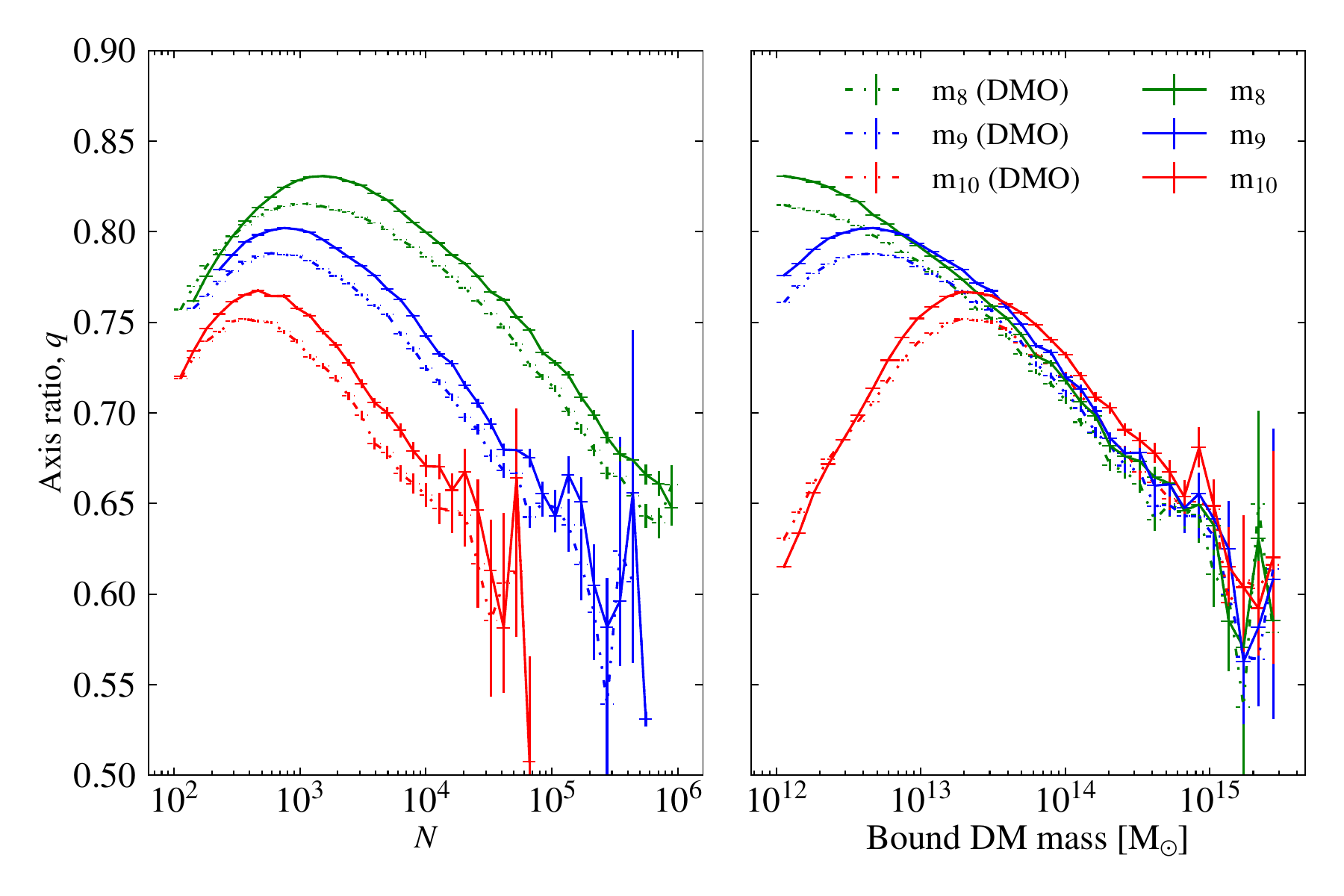}
    \caption{Left: The variation of the average 3D axis ratio, $q = b/a$, with bound dark matter particle number for the three resolutions for both the HYDRO (solid) and DMO (dashed) runs. Right: The variation of the average 3D axis ratio with bound halo dark matter mass for the three resolutions for both the HYDRO and DMO runs. The different colors indicate the different simulation resolutions, with $\mathrm{m}_8$ being the highest resolution. There is a downturn at below $10^3$ particles (which corresponds to a different mass for each resolution as is clear from the right panel) which suggests its origin is numerical rather than physical. The DM masses for the HYDRO runs have been corrected by a factor of $\Omega_\mathrm{m}/(\Omega_\mathrm{m} - \Omega_\mathrm{b})$. The errors are calculated for each particle number (or mass) bin by taking the standard deviation in each bin and dividing by the square root of the number of objects in that bin.}
    \label{fig:q_m_n}
\end{figure*}

\begin{table}
\caption{Simulation details of the runs used in this work.}
\label{table:sim_details}
\centering
\begin{tabular}{ccccrS[table-format=3.2]} 
\hline
\noalign{\smallskip}
Identifier & $\Delta m_*$ & $\Delta f_{\mathrm{gas}}$ & AGN & $N_\mathrm{b}$ & $\mathrm{m}_{\mathrm{CDM}}$\\
 & ($\sigma$) & ($\sigma$) &  & & {($10^{9} \mathrm{M}_{\odot}$)} \\
 \noalign{\smallskip}
  \hline
  L1\_m8 ($\mathrm{m}_8$) & 0 & 0 & thermal & $3600^3$ & 0.706\\
  L1\_m9 ($\mathrm{m}_9$)& 0 & 0 & thermal & $1800^3$ & 5.65\\
  L1\_m10 ($\mathrm{m}_{10}$)& 0 & 0 & thermal & $900^3$ & 45.2\\
  L1\_m8\_DMO & - & - & - & $3600^3$ & 0.84\\
  L1\_m9\_DMO & - & - & - & $1800^3$ & 6.72\\
  L1\_m10\_DMO & - & - & - & $900^3$ & 53.8\\
  L2p8\_m9 & 0 & 0 & thermal & $5040^3$ & 5.65\\
  fgas+2$\sigma$ & 0 & +2 & thermal & $1800^3$ & 5.65\\
  fgas-2$\sigma$ & 0 & -2 & thermal & $1800^3$ & 5.65\\
  fgas-4$\sigma$ & 0 & -4 & thermal & $1800^3$ & 5.65\\
  fgas-8$\sigma$ & 0 & -8 & thermal & $1800^3$ & 5.65\\
  M*-$\sigma$ & -1 & 0 & thermal & $1800^3$ & 5.65 \\
  Jet & 0 & 0 & jets & $1800^3$ & 5.65\\
  \hline
\end{tabular}
\tablefoot{The L2p8\_m9 run has a box size of 2.8 Gpc while the rest have a box size of 1 Gpc. Identifier refers to the simulation run, where m8, m9 and m10 indicate $\log_{10}$ of the mean baryonic particle mass. When absent, a feedback variation at the $\mathrm{m}_9$ resolution is being referred to. $\Delta m_*$ is the number of standard deviations by which the observed stellar masses are shifted before calibration, $\Delta f_{\mathrm{gas}}$ is the number of standard deviations the observed cluster gas fractions are shifted before calibration, AGN is the form of AGN feedback implemented, $N_\mathrm{b}$ is the number of baryonic particles and $\mathrm{m}_{\mathrm{CDM}}$ the mean CDM particle mass.}
\end{table}

Another useful feature of the FLAMINGO simulations are the 1 Gpc runs with identical initial conditions but different particle mass resolutions. The three resolutions are the low ($\mathrm{m}_{\mathrm{CDM}} = 4.52 \times 10^{10} \ \mathrm{M}_{\odot}$), intermediate ($\mathrm{m}_{\mathrm{CDM}} = 5.65 \times 10^9 \ \mathrm{M}_{\odot}$) and high ($\mathrm{m}_{\mathrm{CDM}} = 7.06 \times 10^8 \ \mathrm{M}_{\odot}$) resolution runs, subsequently referred to as the $\mathrm{m}_{10}$, $\mathrm{m}_{9}$ and $\mathrm{m}_{8}$ runs. The $\mathrm{m}_{8}$, $\mathrm{m}_{10}$ and hydro-fiducial $\mathrm{m}_{9}$ runs were re-calibrated to match the same data. We used these to explore resolution effects on the shapes of haloes. All the different baryonic feedback models were run at the $\mathrm{m}_{9}$ resolution. The different feedback modes implemented varied the AGN feedback mechanism between thermal and jet feedback. The sub-grid parameters controlling baryonic feedback were tuned to reproduce the observed redshift zero stellar mass function and gas fractions in galaxy groups and clusters at low redshift. As the observations have uncertainties, several feedback variations were created by calibrating the sub-grid parameters to gas fractions or stellar mass functions that are shifted away from their fiducial values, as described in detail by \citet{Kugel2023}. Each variation was defined with respect to the observable it was calibrated to, that is the number of standard deviations by which the observed stellar masses and cluster gas fractions were shifted prior to calibrating the sub-grid parameters. We also used the 2.8 Gpc box, which was run at the intermediate mass resolution and with the fiducial feedback model, which provided us with unprecedented statistics at the high-mass end.

Halo finders were applied to the simulations to create halo catalogues. Haloes were found using a 3D Friends-of-Friends (FoF) algorithm with linking length $l = 0.2$ times the mean CDM interparticle separation. VELOCIRAPTOR \citep{Elahi2019} was then used to decompose the FoF haloes into self-bound subhaloes that are dynamically distinct from the mean background halo using 6D phase space information. In the case of the alternate substructure finder HBT-HERONS \citep{Han2018, Moreno2025}, haloes were tracked across redshift snapshots to better account for their merging history. A detailed comparison of different halo finders has been presented by \citet{Moreno2025}.

The choice of halo finder could also change which particles are assigned to each halo, thereby changing their shapes. The effect of halo finders on the shapes of haloes is not explored in the literature, even though this could impact results. In most cases described in this paper, we refer to haloes found by VELOCIRAPTOR. We also compared our results with haloes found using HBT-HERONS (see Sect. \ref{sect:methods}).

\subsection{Inertia tensors and shapes}

To estimate the shape of a given halo, we calculated the inertia tensor for each halo. Haloes were modelled as 3D ellipsoids whose axes point in the direction of the eigenvectors of the Simple Inertia Tensor (SIT) defined as:

\begin{equation}
\label{eqn:sit}
    I_{ij} = \frac{1}{M}\sum_n m_{(n)}x_{i}^{(n)}x_{j}^{(n)},
\end{equation}
where $i, j = {1, 2, 3}$ correspond to the three axes of the simulation box, $m_n$ is the mass of the $n$th particle, $x^{(n)}$ are the positions of the $n$th particle in the $i$ or $j$ direction and $M$ is the total mass of the object. 

We used the SIT calculated in a non-iterative scheme for all shapes in this work, centered on the centre of potential (COP; position of the most bound particle). There has been some discussion in the literature regarding the differences between shapes measured via different types of inertia tensors. \citet{Zemp2011} advocated an iterative scheme where the ellipsoidal shell around particles within which the inertia tensor is calculated is updated to match the shape found in the previous iteration. \citet{Valenzuela2024} found that unweighted and reduced methods led to systematic biases in the shapes. On the other hand, \citet{Bett2012} found little difference between the iterative and non-iterative methods for the SIT, as did \citet{Velliscig2015a} (see their Appendix A1). There is no consensus in the literature at the moment regarding which inertia tensor scheme is best, which can also differ depending on the application. 

Since we were interested in how well the satellite galaxies traced the shape of their host dark matter halo, the SIT is most appropriate from an observational perspective as satellites are equally weighted. Moreover, we found little difference in the distribution of shapes of the host dark matter haloes with different methods in the mass range probed here, so we opted to use the SIT. We leave a more comprehensive comparison to future work.

The eigenvalues of the inertia tensor in Eqn. \ref{eqn:sit} are denoted by $\lambda_i$, where $i = {1, 2, 3}$ correspond to the three axes of the ellipsoid. The lengths of the three axes of this ellipsoid, the major, intermediate and minor axes, are given by $a = \sqrt{\lambda_1}$, $b = \sqrt{\lambda_2}$ and $c = \sqrt{\lambda_3}$ respectively. We further define the axis ratio $q$, sphericity $s$ and triaxiality $t$ as

\begin{align}
    q &= b/a \\
    s &= c/a \\
    t &= (a^2 - b^2)/(a^2 - c^2)
\end{align}
The eigenvectors of the inertia tensor $\hat{e}_i$ encode the orientation of the structure itself.

\begin{figure*}

    \centering
    \includegraphics[scale=0.5]{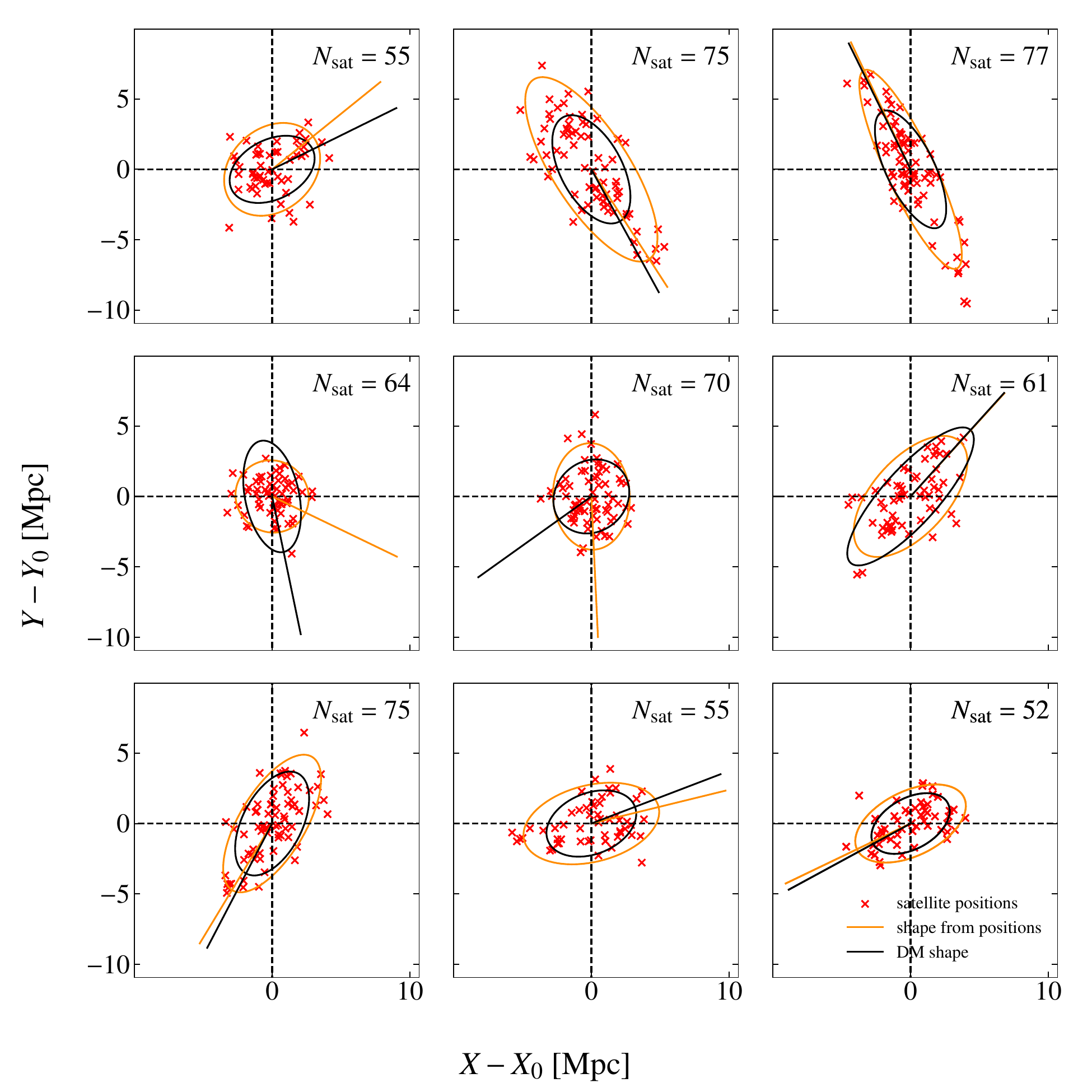}
    \caption{Examples of haloes from the $\mathrm{m}_9$ simulation with more than 50 satellite galaxies (with bound dark matter mass in the range 1.3 to 2.4 $\times 10^{15} \ \mathrm{M}_{\odot}$). The red crosses denote the positions of galaxies with the ellipse calculated using these positions shown in orange. The black ellipse is the corresponding dark matter ellipse. The lines show the orientation angles of the ellipses. All ellipses shown are derived from projected shapes. The positions are centred on the Centre of Potential (COP, position of the most bound particle) of the (sub)halo.}
\label{fig:satellite_shapes}

\end{figure*}

\subsection{The impact of noise}

The axis lengths $a$, $b$ and $c$ are noisy quantities, and ratios of noisy quantities are biased, such as ellipticities derived from noisy galaxy light profiles \citep{Melchior2012}. Similarly, inertia tensors calculated with very few particles result in biased axis ratios. In simulations, this corresponds to having too few particles for a given halo, and in observations, too few satellite galaxies per host. In principle, if the bias arises purely from sampling noise, we can predict this effect and thus statistically correct the shape measurement of haloes. 

\begin{figure*}
    \centering
    \includegraphics[scale=0.45]{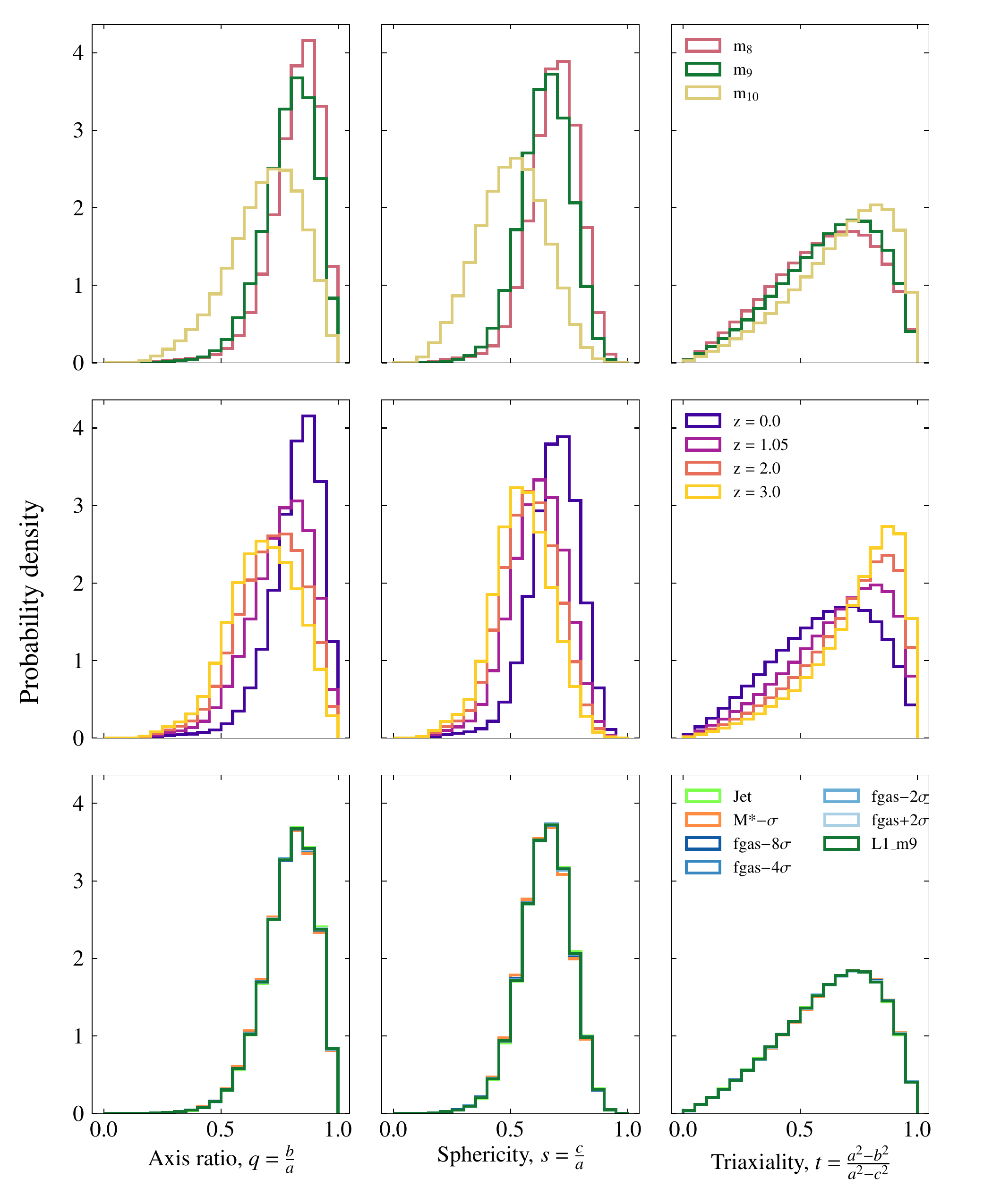}
    \caption{The distributions of 3D axis ratio ($q$), sphericity ($s$) and triaxiality ($t$) for the dark matter of haloes with dark matter mass (bound) more than $10^{12}$ $\mathrm{M}_{\odot}$; Top panel: for the three resolutions of the simulation used, Middle panel: for the highest resolution run used in this work ($\mathrm{m}_8$) across redshift, Bottom panel: for the different feedback models implemented at intermediate resolution. Shapes from the $\mathrm{m}_9$ are reasonably well converged, whereas the $\mathrm{m}_{10}$ is less reliable. Halo shapes vary significantly with redshift. There is no strong variation with feedback variation. Mean values are shown in Table \ref{table:mean_axis_ratio}.}
    \label{fig:q_s_t}
\end{figure*}

To understand how the number of particles affects the shape measurement, we made use of the three different resolution runs with identical initial conditions and similar calibration to data, so we could study the effects of resolution on the shapes of haloes. Since we are interested in the variation of properties with halo particle number, we need to choose an appropriate mass definition. We use the bound mass of the dark matter haloes instead of other mass definitions (namely $\mathrm{M}_{200\mathrm{crit}}$) as this is the mass of all the particles assigned to a particular halo by the halo finder. Previous work has shown that the shape of a halo depends on its mass \citep{Jing2002, Allgood2006, Despali2014}. We therefore explored the dependence of the 3-D axis ratio on the dark matter mass of the halo, which is shown in Fig.\ref{fig:q_m_n}. We plot the axis ratios of the dark matter haloes of host haloes as a function of particle number (left) and total halo dark matter mass (right) for both dark matter only (DMO) and hydrodynamics (HYDRO) runs at the three different resolutions. A halo in the DMO run has more dark matter mass than its equivalent in the HYDRO run. In order to compensate for this and make a one-to-one mass comparison between these runs, we increased the HYDRO dark matter masses by a factor of $\Omega_\mathrm{m}/(\Omega_\mathrm{m} - \Omega_\mathrm{b})$ where $\Omega_\mathrm{m}$ is the the total matter density parameter and $\Omega_\mathrm{b}$ is the universal baryonic matter density. 

Consistent with previous work \citep[e.g.][]{Jing2002, Jeeson-Daniel2011, Schneider2012, Bryan2013, Despali2014, Velliscig2015a}, we find that lower mass haloes have higher axis ratios (i.e., they are rounder), for both the hydrodynamical and gravity-only runs. Moreover, shapes resulting from pure dark matter dynamics are different than when hydrodynamics are included. As the left panel of Fig.\ref{fig:q_m_n} shows, there is a downturn in the axis ratios of haloes containing fewer than about $10^3$ particles, which is due to resolution effects. Below this limit, we cannot trust the shapes obtained from the highest resolution run considered in this work (i.e., the $\mathrm{m}_8$ run). Therefore, we confined the analysis to haloes with dark matter mass $> 10^{12} \ \mathrm{M}_{\odot}$.  The 3D axis ratios of the $\mathrm{m}_9$ and $\mathrm{m}_{10}$ runs have not converged with the $\mathrm{m}_8$ run at bound dark matter masses $\sim 10^{12} - 10^{14} \ \mathrm{M}_{\odot}$. Since we are interested in quantifying these resolution effects, we employed the same mass cut for each resolution run, which thus corresponds to a different minimum number of particles. 

As dynamical friction causes higher mass haloes to be closer to the halo center, and observational surveys may not be deep enough to capture low mass satellites, the mass cut should be varied to test whether our results are robust to such effects. We explore the effect of changing this mass cut in Appendix \ref{sect:appendixA}.

We wish to investigate how well the shape of a dark matter halo is traced by its satellite galaxies, which is illustrated in Fig.\ref{fig:satellite_shapes}. Examples of clusters from the simulation with more than 50 satellite galaxies (with bound dark matter mass $> 10^{12} \ \mathrm{M}_{\odot}$) are shown with the red crosses indicating satellite positions, centred on the centre of mass determined using their positions. The example host haloes have a bound dark matter mass between 1.3 - 2.4 $\times 10^{15} \ \mathrm{M}_{\odot}$. A satellite is defined as a subhalo found by VR which has a host, and we use the centre of potential of this subhalo as its position. The ellipses derived from satellite positions are shown in orange, and the black ellipses are the corresponding host halo dark matter shapes. We have considered only projected axis ratios here. When deriving shapes from the satellite positions, we use Eqn. \ref{eqn:sit} with each position being assigned equal mass (i.e. no mass-weighting). The orientation angle of each ellipse is also shown, with the misalignment angle defined as the angle between the orange and the black lines. We can see that the shapes and orientation angles for the two methods generally match very well with each other, with the dark matter haloes being generally more compact than the sizes derived from the satellite positions. This is due to the fact that the central galaxy contains a lot of mass that makes the eigenvector values of the SIT smaller, whereas it does not contribute very much to the size derived from the satellite positions due to the difference in mass weighting. 

\subsection{Halo shapes}

The distributions of the dark matter shape parameters $q$, $s$ and $t$ for central haloes are shown in Fig.\ref{fig:q_s_t}, with the corresponding mean values listed in Table \ref{table:mean_axis_ratio}. The top panel of Fig.\ref{fig:q_s_t} compares the distribution of axis ratios for the three different particle resolutions. We observe that the shapes from the $\mathrm{m}_{9}$ run are reasonably well converged, whereas the $\mathrm{m}_{10}$ results are not. Lower resolution leads to more elliptical haloes (i.e., lower $q$ and $s$, and higher $t$). 

The middle panel of Fig.\ref{fig:q_s_t} depicts the distribution of the shape parameters for the $\mathrm{m}_8$ run, at different redshifts. We considered clusters in the same run at redshift snapshots of $z = 0, 1.05, 2$ and 3. The most notable variation with redshift is that of $t$, which increases with redshift for haloes. $t \le 1/3$ indicates an oblate ellipsoid, $t \ge 2/3$ indicates a prolate ellipsoid, and $1/3 \le t \le 2/3$ are triaxial ellipsoids \citep{Schneider2012, Valenzuela2024}. As already noted by \citet{Schneider2012}, haloes become more prolate with mass and redshift. The shapes of haloes are nearly spherical as $q$ is close to 1, consistent with previous results showing moderate ellipticity for haloes at these masses. The bottom panel of Fig.\ref{fig:q_s_t} shows the variation of shapes for the different feedback models of the $\mathrm{m}_9$ run. There is no noticeable change in the distributions of shapes for central dark matter haloes above $10^{12} \ \mathrm{M}_{\odot}$, despite the wide range of baryonic feedback scenarios adopted.

\begin{table}
\caption{Mean values of axis ratios, sphericity and triaxiality}
\label{table:mean_axis_ratio}
\centering
\begin{tabular}{cccccc} 
\hline
\noalign{\smallskip}
 Resolution & Feedback mode & $z$ & $\langle q \rangle$ & $\langle s \rangle$ & $\langle t \rangle$\\
 \noalign{\smallskip}
  \hline
  \noalign{\smallskip}
  $\mathrm{m}_8$ & fiducial & 0 & 0.82 & 0.68 & 0.60 \\
  $\mathrm{m}_9$ & fiducial & 0 & 0.79 & 0.65 & 0.62\\
  $\mathrm{m}_{10}$ & fiducial & 0 & 0.69 & 0.50 & 0.68 \\
  
  $\mathrm{m}_8$ & fiducial & 1.05 & 0.75 & 0.61 & 0.66 \\
  $\mathrm{m}_8$ & fiducial & 2.0 & 0.71 & 0.58 & 0.71 \\
  $\mathrm{m}_8$ & fiducial & 3.0 & 0.68 & 0.55 & 0.74\\
  
  $\mathrm{m}_9$ & fgas+2$\sigma$ & 0 & 0.79 & 0.65 & 0.62 \\
  $\mathrm{m}_9$ & fgas-2$\sigma$  & 0 & 0.79 & 0.65 & 0.62 \\
  $\mathrm{m}_9$ & fgas-4$\sigma$ & 0 & 0.79 & 0.65 & 0.62 \\
  $\mathrm{m}_9$ & fgas-8$\sigma$ & 0 & 0.79 & 0.65 & 0.62 \\
  $\mathrm{m}_9$ & M*-$\sigma$ & 0 & 0.79 & 0.65 & 0.63 \\
  $\mathrm{m}_9$ & Jet & 0 & 0.79 & 0.65 & 0.62 \\

  \hline
\end{tabular}
\tablefoot{Shown are mean values across resolution, redshift and feedback modes for haloes with bound dark matter mass $> 10^{12} \ \mathrm{M}_{\odot}$.}
\end{table}

\begin{figure}
    \centering
    \includegraphics[scale=0.5]{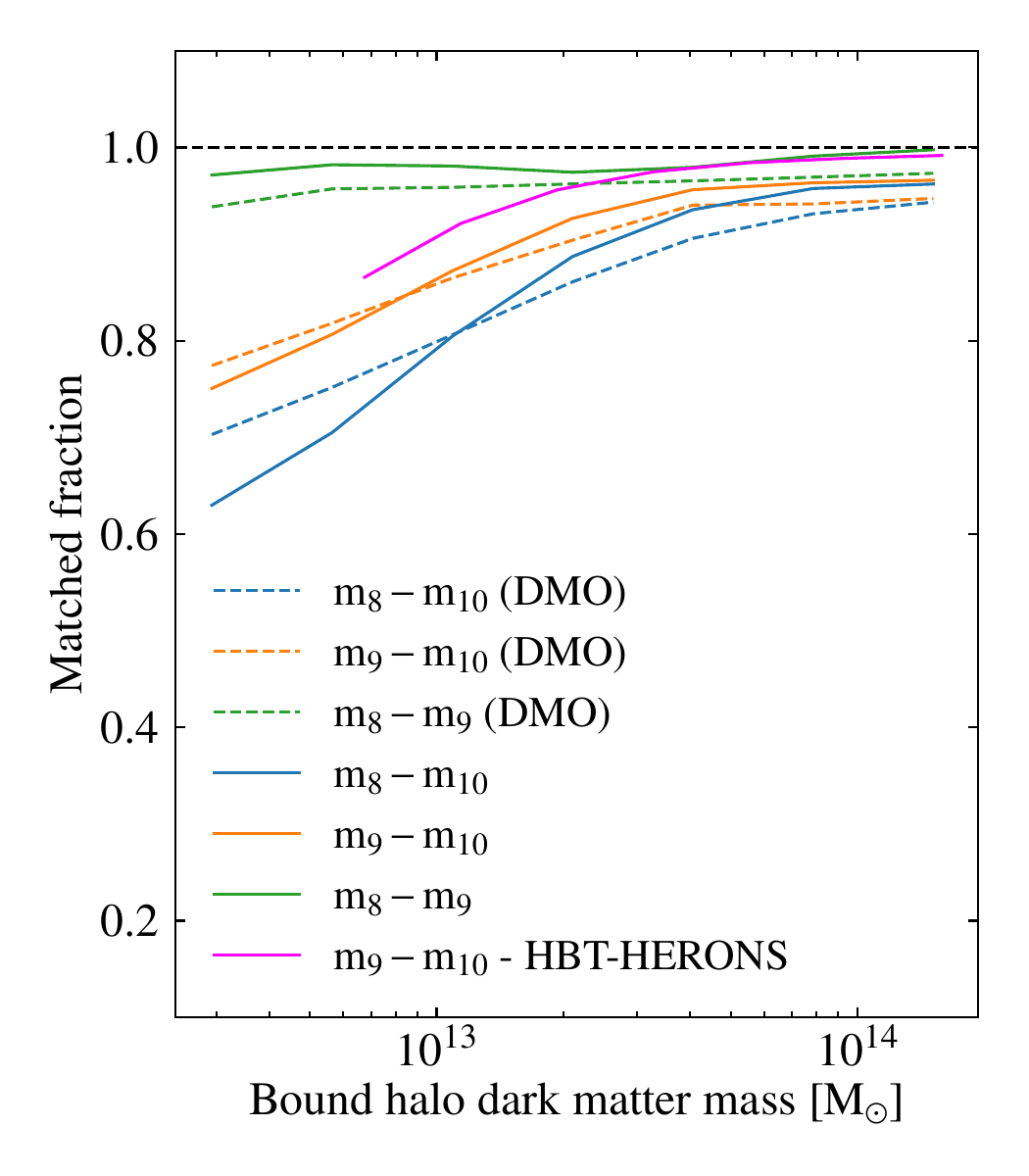}
    \caption{Fraction of haloes matched across different mass resolution runs as a function of halo dark matter mass. The dashed lines indicate DMO runs whereas the solid lines are for HYDRO. The Magenta line is for the HBT-HERONS halo finder.}
    \label{fig:halo-matching}
\end{figure}

\subsection{Halo Matching}

In order to test the validity of our approach, we also estimated the axis ratio bias using the $\mathrm{m}_8$, $\mathrm{m}_9$ and $\mathrm{m}_{10}$ runs. Before comparing the shapes of haloes at different mass resolutions, we match them so that the same objects are being compared. Much of the literature has focused on halo matching at the same resolution, typically between DMO and hydro runs, or with different baryonic feedback models. We used a similar approach, where we matched haloes at the same positions in the box, with similar masses. 

When matching between two resolutions, we searched for a halo in the lower resolution run at a position within 1 Mpc of the position in the higher resolution run. We used the COP as a proxy for the position of the halo. If a match was found, we checked that the masses of the two haloes are within 30 per cent of each other. 

An interesting problem arises when matching haloes across resolutions. It is possible that a low resolution halo corresponds to multiple separate objects at higher resolution. These objects will obviously have different shapes, and as it is our goal to measure the shape error due to having different numbers of particles in a halo, this scenario is not of interest to us. Thus, we excluded these multiply matched haloes from the analysis. Correcting for these objects (which made up only 5 per cent of cluster haloes at the high mass end) we arrived at a 99.75 per cent matched fraction for the $\mathrm{m}_8$-$\mathrm{m}_9$ matching at the high mass end (shown in Fig.\ref{fig:halo-matching}). We checked that the excluded haloes do not have a different distribution of shapes that could have lead to possible selection biases. We can also see that the matched fraction for the HBT-HERONS halo finder is consistently higher at all masses when comparing with VR for the $\mathrm{m}_9$-$\mathrm{m}_{10}$ case. 

\section{Determination of sampling bias}
\label{sect:methods}

Although the different simulation runs can provide an estimate of the axis ratio bias, it is difficult to determine the source of the bias. In the simulation, in addition to the sampling bias, unresolved dynamics and sub-grid effects could potentially create biases in the shape. In order to estimate the bias for the inertia tensor shapes, we employed a Monte Carlo (MC) setup that captures only the sampling noise, and a comparison with the simulation will serve to illustrate how much of the bias in the simulations can be explained by sampling effects. The details of our MC setup is described below.

For a specific halo, the bias from sampling noise depends on its radial profile. A halo of a given shape and density profile is created with a specific number of particles, $N$. The inertia tensor was calculated using this ensemble of particles, and the resulting shape was compared with the input shape. Since we can only observe projected shapes, we focused our results on the difference between the input 2D axis ratio and the derived 2D axis ratio. We quantify this by the axis ratio bias, defined as $(q_{\mathrm{m}} - q_{\mathrm{t}})/{q_{\mathrm{t}}}$ where $q_{\mathrm{m}}$ is the measured axis ratio and $q_{\mathrm{t}}$ is the true axis ratio. 

The MC haloes are ellipsoidal, with the density following a Navarro-Frenk-White profile \citep{Navarro1996} 
\begin{equation}
    \rho(r) = \frac{\delta_c \ \rho_{\mathrm{crit}}}{r/r_{\mathrm{s}}(1+r/r_{\mathrm{s}})^2},
\end{equation}

\begin{figure}
    \centering
    \includegraphics[scale=0.6]{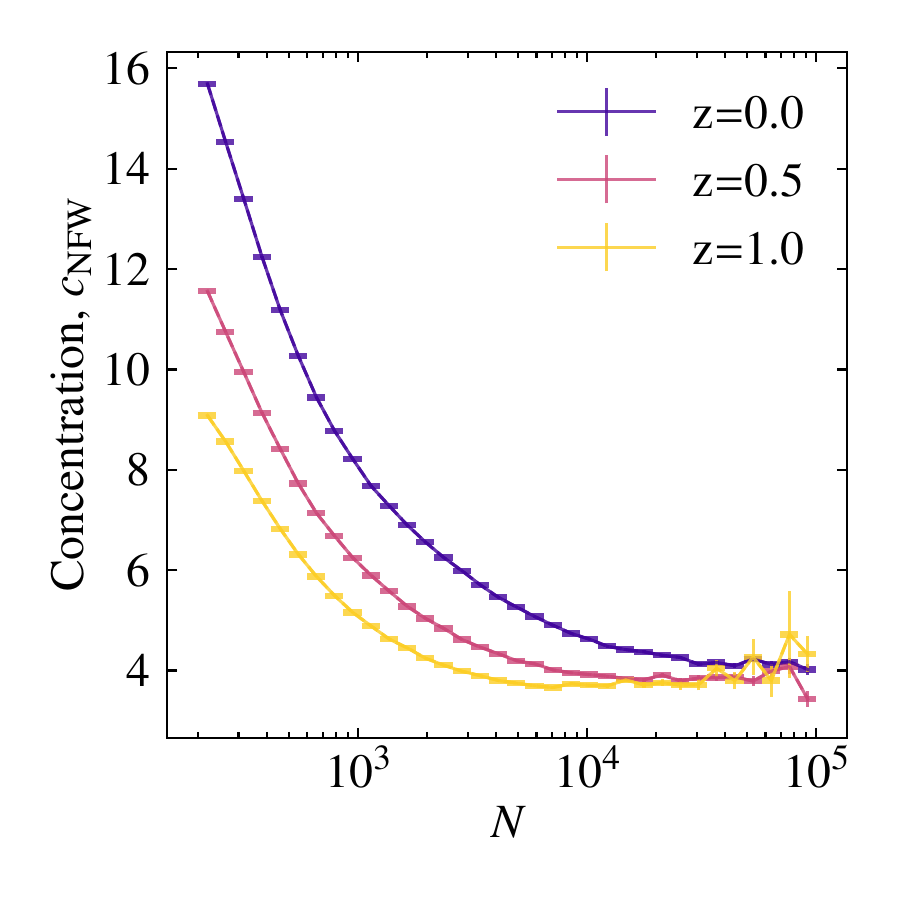}
    \caption{Variation of total matter concentration with dark matter particle number for the $\mathrm{m}_9$ HYDRO run at different redshifts. At each $N$ and at a particular redshift, the corresponding concentration is used to create mock NFW haloes in the MC exercise. These curves are used down to $N$ = 180, which is sufficient as we only considered haloes with a mass well above this particle number limit with the $\mathrm{m}_9$ resolution.}
    \label{fig:m9_N_concentration}
\end{figure}

\begin{figure}
    \centering
    \includegraphics[scale=0.5]{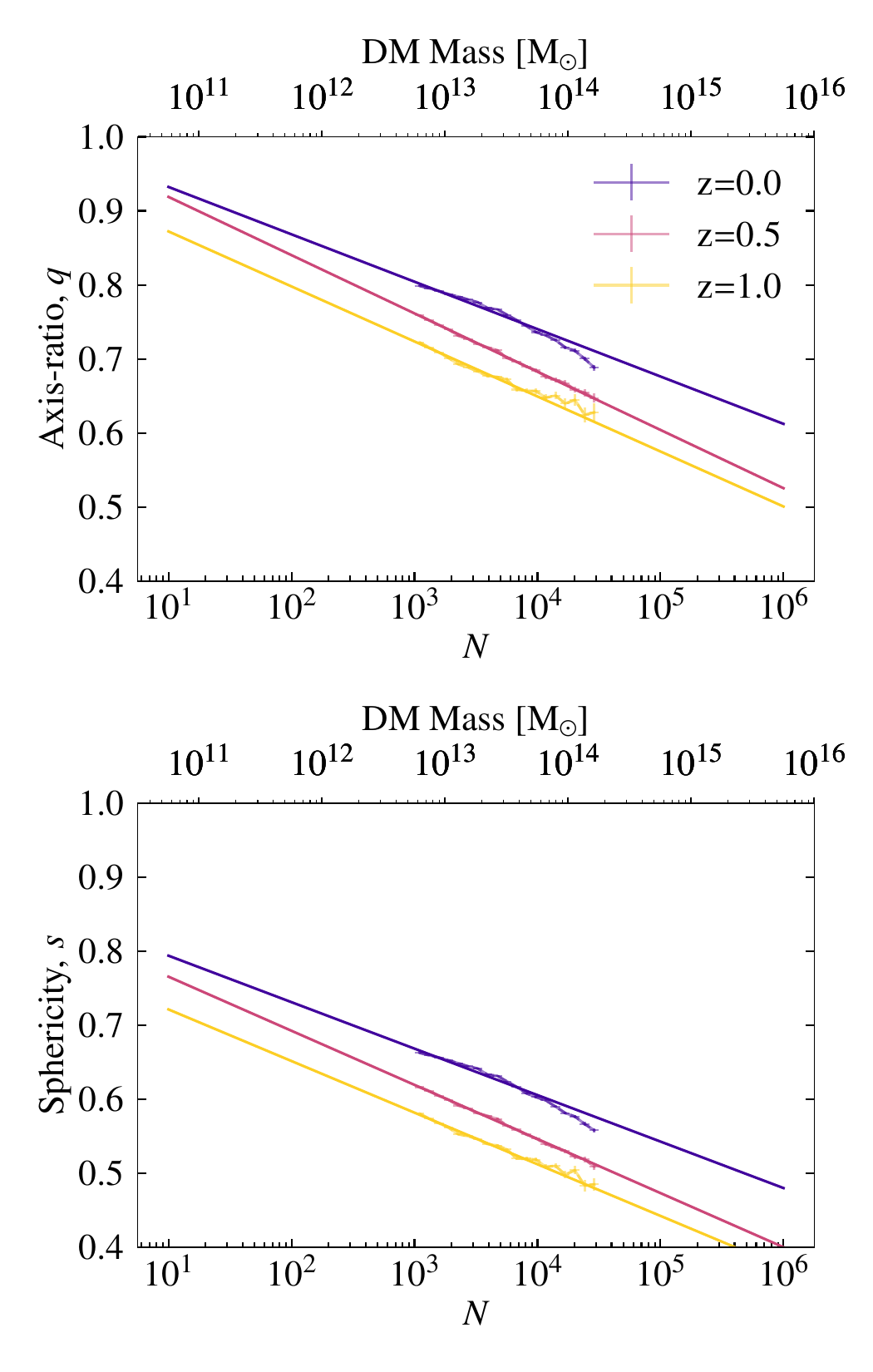}
    \caption{Variation of the mean $q$ and $s$ with dark matter particle number for the $\mathrm{m}_9$ HYDRO run at different redshifts. The corresponding fits are used to sample the shapes of ellipsoidal NFWs in the MC exercise.}
    \label{fig:m9_q_s_N_fit}
\end{figure}

\begin{figure*}
  \centering
  \subfloat[a][]{\includegraphics[scale=0.42]{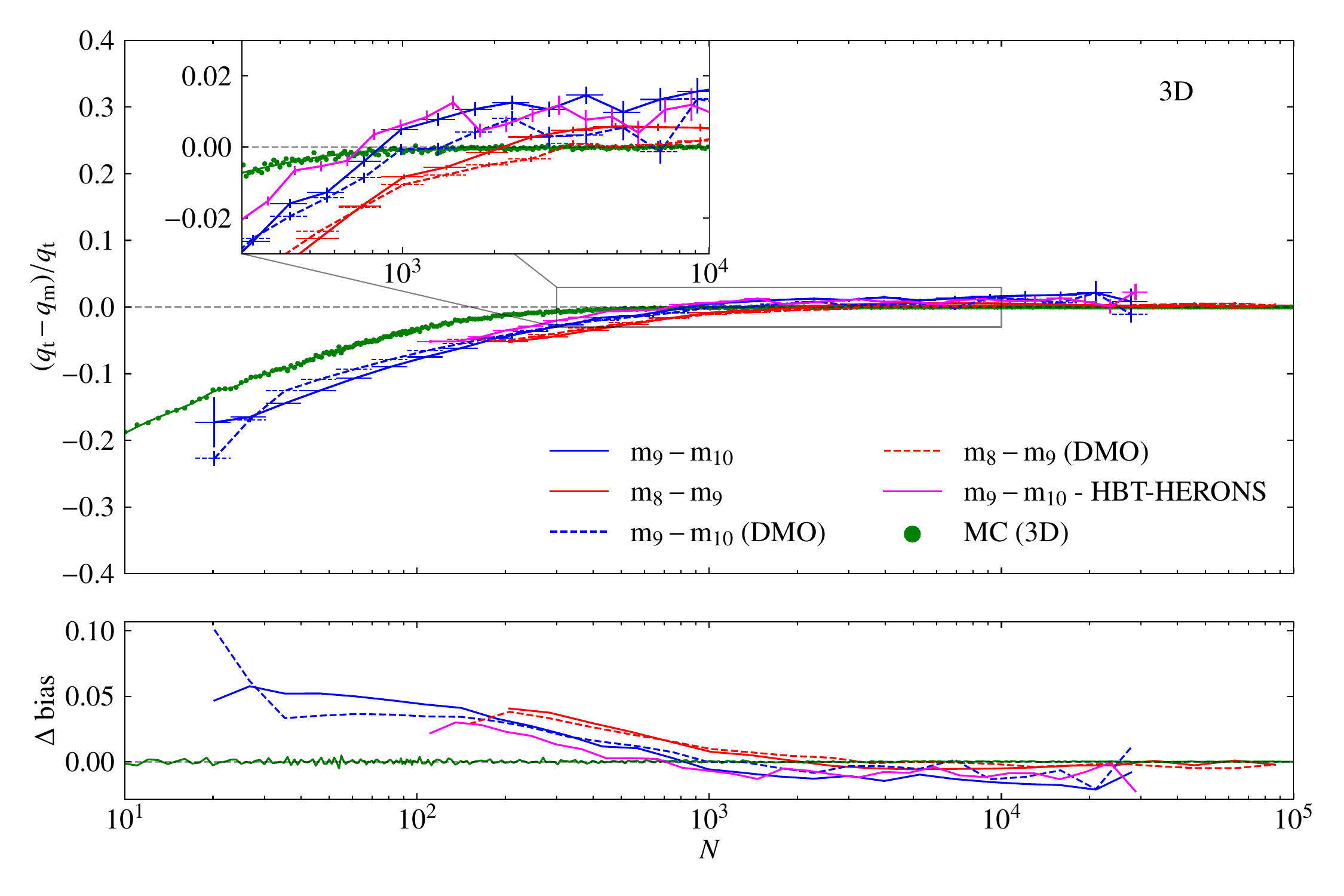} \label{fig:a}} \\
  \subfloat[b][]{\includegraphics[scale=0.42]{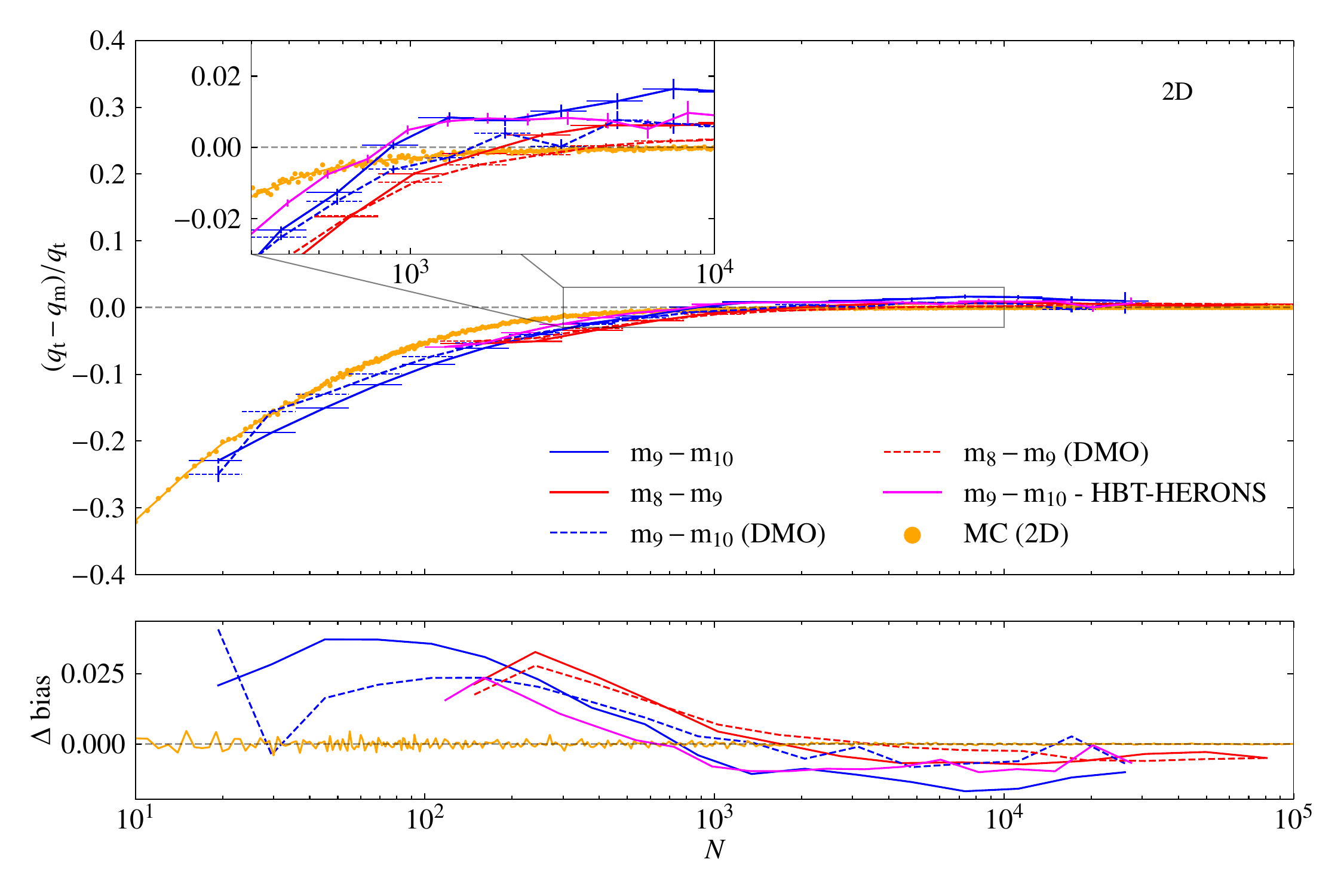} \label{fig:b}}
  \caption{Variation of median axis ratio bias with halo dark matter particle number for (a) 3-D and (b) 2-D (projection). Axis ratio bias is defined as $(q_{\mathrm{t}}-q_{\mathrm{m}})/q_{\mathrm{t}}$ where $q_{\mathrm{t}}$ is the true axis ratio and $q_{\mathrm{m}}$ is the measured axis ratio. The bias estimate assuming sampling noise from the MC is shown in green in sub-figure (a) and yellow in sub-figure (b). This is compared with the bias calculated between haloes matched between different resolution FLAMINGO simulations ($\mathrm{m}_8$, $\mathrm{m}_9$ and $\mathrm{m}_{10}$) for both HYDRO (solid lines) and DMO (dashed lines) in the bottom panels of each sub-figure. $\Delta \ \mathrm{bias}$ is the difference between the bias estimated with the MC and that from the simulation. The errors for each bin in $N$ were calculated by taking the standard deviation in each bin and dividing by the square root of the number of objects in that bin. For clarity, the 1$\sigma$ region for the bias from the MC is not plotted here. VR is used for all cases, except the magenta lines which use HBT-HERONS.}
  \label{fig:q_bias}  
\end{figure*}

\begin{figure}
    \centering
    \includegraphics[scale=0.5]{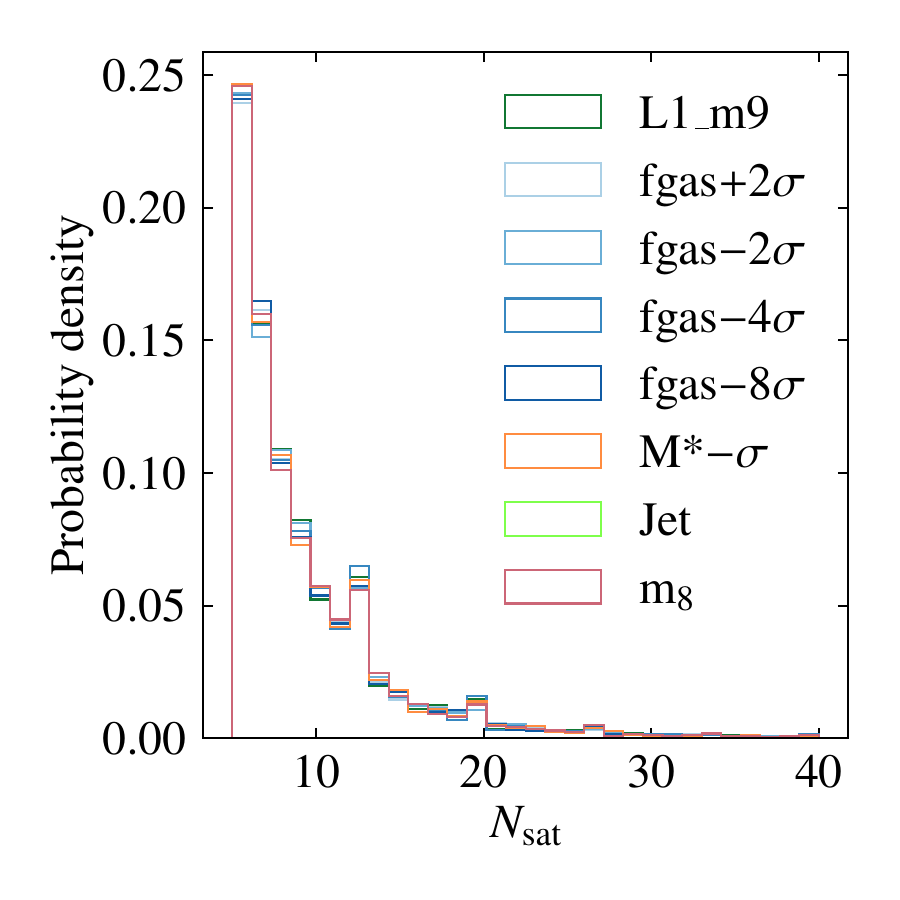}
    \caption{Probability distribution of number of satellites more massive than $10^{12} \ \mathrm{M}_{\odot}$ (bound) in each halo for the $\mathrm{m}_8$ run and the feedback variations of the $\mathrm{m}_9$ run.}
    \label{fig:satellite_numbers}
\end{figure}

\noindent where $r_{\mathrm{s}}$ is a scale radius, $\delta_c$ is a characteristic density and $\rho_{\mathrm{crit}}$ is the critical density \citep{Navarro1997}. Halo concentration is described as $c_{\mathrm{NFW}} \equiv r_{\mathrm{vir}}/r_{\mathrm{s}}$ where $r_{\mathrm{vir}}$ is the virial radius (here we have used $R_{200\mathrm{crit}}$).\footnote{We use the subscript NFW to distinguish concentration from the axis-length $c$.} The concentration of a halo is related to its mass, and we use a mass-concentration relation derived from the $\mathrm{m}_9$ run to create realistic NFW haloes. The variation of concentration with particle number (for the $\mathrm{m}_9$ particle resolution) and redshift is shown in Fig.\ref{fig:m9_N_concentration}. We used concentrations calculated for all particles within $R_{200\mathrm{crit}}$ from the centre of the halo. These values were calculated using a similar approach to that detailed in \citet{Wang2024}, and limited to central haloes. Since we do not have concentrations for satellite galaxies, we only have a sufficient number of haloes with a reasonable scatter on the concentration value in low $N$ bins down to about 180 particles per halo. This is sufficient for our purposes as we considered host haloes with at least 5 satellite galaxies above the mass cut of $10^{12} \ \mathrm{M}_{\odot}$, which is well above the mass corresponding to $N = 180$ particles at the $\mathrm{m}_9$ particle resolution. We also created mock haloes where concentration values were randomly sampled for each halo (i.e., with no relation between the mass and the concentration of the halo) and found that this does not affect the results. We thus conclude that our results are insensitive to the assumed concentrations and leave a more sophisticated treatment of the mass-concentration relation to future work.

\begin{figure*}
    \centering
    \includegraphics[scale=0.39]{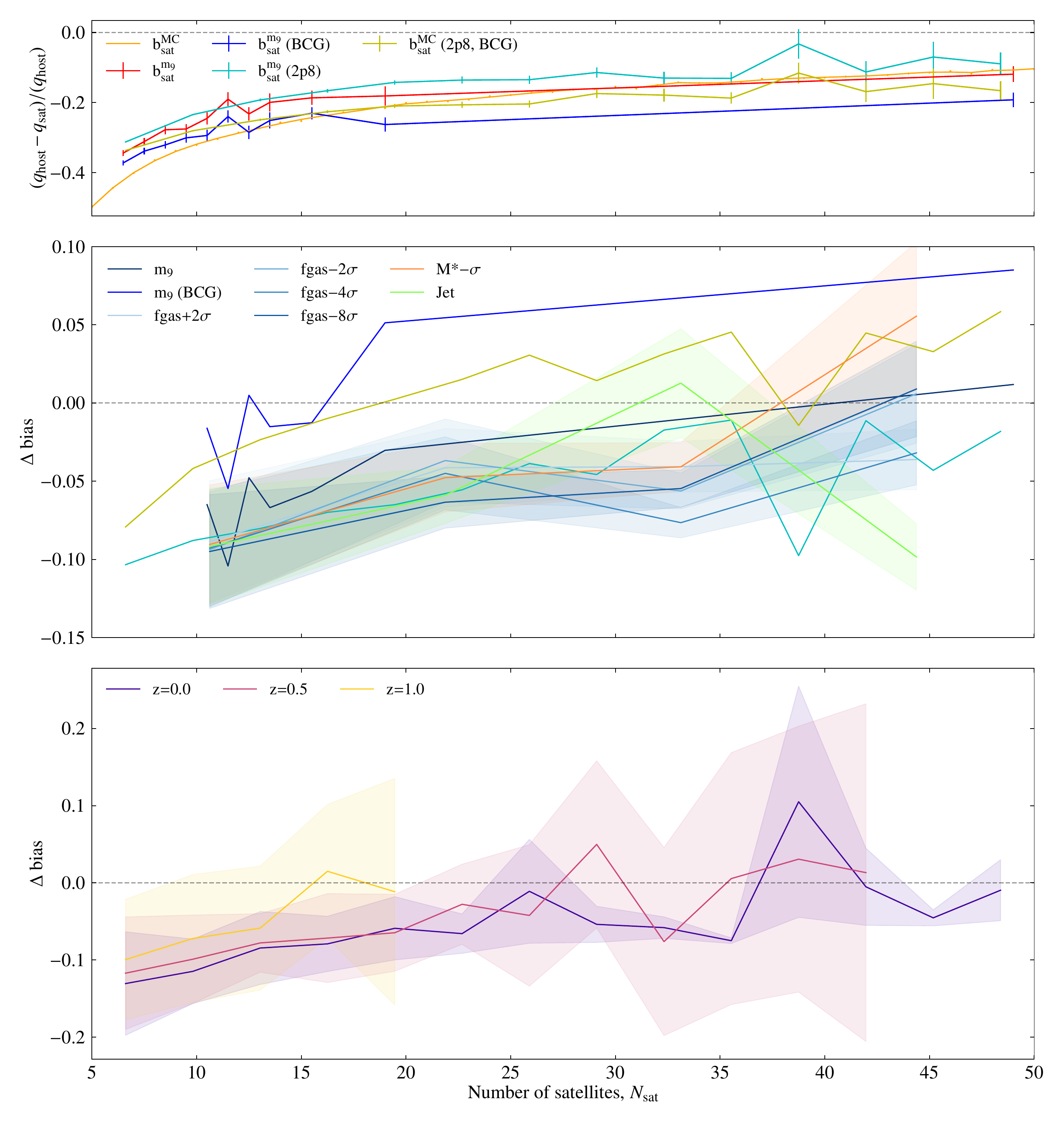}
    \caption{Top panel: Variation of the median projected axis ratio bias with number of satellites. We compare the bias estimate from the MC with the actual bias from measuring shapes with satellites in the $\mathrm{m}_9$ case. Here the bias is defined as $(q_{\mathrm{host}} - q_{\mathrm{sat}})/q_{\mathrm{host}}$ where $q_{\mathrm{host}}$ is the axis ratio of the host DM halo and $q_{\mathrm{sat}}$ is that measured using the distribution of satellite galaxies. ${\mathrm{b}}^{\mathrm{m}_9}_{\mathrm{sat}}$ denotes the bias when using the satellite positions to estimate the shape of the halo for the $\mathrm{m}_9$ run, whereas ${\mathrm{b}}^{\mathrm{m}_9}_{\mathrm{sat}}$ (BCG) denotes the case where the BCG is assumed to be the centre of the shape. 2p8 refers to the 2.8 Gpc box. The middle panel shows the deviation of the estimated bias from the MC exercise for the different hydrodynamic variations. $\Delta \ \mathrm{bias} = \mathrm{b}_{\mathrm{MC}} - \mathrm{b}_{\mathrm{sat}}$, and the error bars (shown by the shaded regions) are estimated using the jack-knife technique using 27 sub-boxes. The bottom panel shows the deviation from the MC exercise at different redshifts for the 2.8 Gpc run. The error bars are estimated by averaging results from 8 sub-boxes.}
    \label{fig:sat_number-q_bias}
\end{figure*}

As there is also a relation between the mass of a halo and its shape (shown in Fig.\ref{fig:q_m_n}), haloes with more particles have a different intrinsic shape compared to those with fewer particles at a fixed resolution. In order to estimate reasonable values of the axis ratio and sphericity, we fit a relation for these quantities using the $\mathrm{m}_9$ run. This is shown in Fig.\ref{fig:m9_q_s_N_fit}. We only used haloes with $10^{3}$-$10^{4.5}$ particles for the fit to ensure the shapes are sufficiently sampled  (as informed by the left panel of Fig.\ref{fig:q_m_n}) and that there are enough haloes to measure a shape with low scatter. The observed relation is well described by a straight line, so we assumed a linear relation in log($N$), and the best-fit parameters are listed in Table \ref{table:fits}.

In order to compare the resulting bias measurements of the MC exercise with the FLAMINGO simulation, we first created a halo with 8 times more particles than required, and then downsampled it to the required number of particles, as the simulation runs differ by factors of 8 in particle numbers. This is different from the approach typically used in the literature to estimate shape convergence with the number of particles, where the same set of simulation haloes are downsampled and a minimum number of particles is chosen for the haloes to be included in the analysis \citep[e.g.][]{Chisari2015, Chisari2017}. 

The same particle distributions were used for both the 3D and 2D cases, with the 2D version being the 3D case projected along the z-axis. We repeated this MC exercise about 5000 times for each value of $N$ in order to estimate the variance within the MC itself.

We tested the validity of the estimate from our MC exercise with haloes from the FLAMINGO run. The variation of the median axis ratio bias with particle number is shown in the top panel of Fig.\ref{fig:q_bias} for the 3D axis ratio and in the bottom panel for the projected case. Haloes matched between resolutions were used to calculate the bias in the axis ratio at different particle numbers from the FLAMINGO simulation, which is then compared to the results of our MC exercise. We calculate $\Delta \ \mathrm{bias}$, the difference between the bias estimated with our MC and the bias from the simulations.
The approximation of pure sampling noise results in a $\Delta \ \mathrm{bias}$ of 0.05 -- 0.1 at the low mass end in 3D and 0.02 -- 0.04 in projection. The bias is estimated better in the case of projected shapes, which is more relevant to observational analyses. $\Delta \ \mathrm{bias}$ also quickly approaches 0 as the number of particles is increased. 

Moreover, we see that our MC matches better with the DMO runs for each combination of resolutions, while the difference between the 2D HYDRO and DMO case is lower the higher the resolution of the runs being compared (i.e., the $\mathrm{m}_8$ - $\mathrm{m}_9$  HYDRO and DMO runs are more similar than for the $\mathrm{m}_9$ - $\mathrm{m}_{10}$ case). We suspect that deviations of the haloes from an NFW profile and other higher order effects would need to be incorporated in the MC to improve these results. We also see that the deviation is marginally lower in the case of the HBT-HERONS halo finder.

We find that a fourth order polynomial captures the axis ratio bias (in projection) shown in the bottom panel of Fig.\ref{fig:q_bias}. This function is given by:

\begin{equation}
\label{eqn:q_bias}
\begin{split}
    \frac{q_{\mathrm{m}} - q_{\mathrm{t}}}{q_{\mathrm{t}}} = -0.005 ({\log_{10}N})^4 + 0.075 ({\log_{10}N})^3 \\ -0.439 ({\log_{10}N})^2 + 1.138 {\log_{10}N} -1.094.
\end{split}
\end{equation}
For a given number of dark matter particles in an ensemble of haloes, $N$, Eqn. \ref{eqn:q_bias} can be used to estimate the level of median projected axis ratio bias from sampling noise.

\begin{table}
\caption{Best fit parameters for the relation between the mean axis ratio $q$ and sphericity $s$ with dark matter particle number $N$.}
\label{table:fits}
\centering
\begin{tabular}{ccccc} 
\hline
\noalign{\smallskip}
 $z$ & $a_q$ & $b_q$ & $a_s$ & $b_s$ \\
 \noalign{\smallskip}
  \hline
  \noalign{\smallskip}
  0  & -0.06401 & 0.99623 & -0.06277 & 0.85669 \\
  0.5 & -0.07870 & 0.99750 & -0.07308 & 0.83867 \\
  1.0 & -0.07431 & 0.94647 & -0.06972 & 0.79107 \\
  \hline
\end{tabular}
\tablefoot{Listed are best fit values at different redshifts for the $\mathrm{m}_9$ run. The functional form of the relations are $q = a_q \log N + b_q$ and $s = a_s \log N + b_s$.}
\end{table}

\section{Axis ratio bias from satellite positions}
\label{sect:results}

In this section, we explore how well satellite galaxies can be used to trace the shape of the underlying dark matter halo. We will study the impact of the choice of halo centre, as well as the dependence on feedback and redshift. In Section \ref{sect:comparison}, we will compare our results with the literature.

\subsection{Satellite galaxies as tracers of halo shape}

To estimate the bias from using satellite galaxies to trace dark matter halo shapes, we compared the axis ratios derived from the satellite positions with those derived from the dark matter particles. We first checked for convergence in the number of satellites hosted by a halo, as this may also depend on resolution, as well as the feedback model implemented. Fig.\ref{fig:satellite_numbers} shows the number of satellites per FoF halo down to a total (sub)halo dark matter mass of $10^{12} \ \mathrm{M}_{\odot}$. For our sample, we only considered haloes with at least 5 satellite galaxies. The number of satellites per halo for the different feedback modes, also shown in Fig.\ref{fig:satellite_numbers}, are very similar to the highest resolution run (shown in black), implying that these statistics have converged, and are robust to feedback (at the mass cut considered here).

First, we must choose a method to determine the centre of the ellipse. One way to do so, often used when estimating shapes of haloes observationally \citep{Zhou2023}, is to assume the position of the BCG as the centre of the halo. We tested this assumption by calculating the distance between the COP of the BCG and the host halo. For each halo above a total dark matter mass of $10^{12} \ \mathrm{M}_{\odot}$, we picked the galaxy with the highest stellar mass as the BCG. We found that in the majority of cases (80 per cent), the central galaxy also contains the most stellar mass, and by definition the central's COP and the host halo's COP coincide. For some cases, however, this is a source of miscentering \citep{Lauer2014}. Another method to infer the centre of the ellipse is to calculate the centre of mass using the positions of the satellites, which is also what we adopted for the MC exercise. Here, each satellite is weighted the same, irrespective of its mass. In the analysis to follow, we will compare the bias from the MC and the simulation, as well as the two methods of determining the centre.

The median bias coming from using the satellite distribution to estimate the halo axis ratio (in projection) is shown in the top panel of Fig.\ref{fig:sat_number-q_bias}. We did not include the BCG in this calculation; both the halo shape and centre were estimated purely from the positions of satellite galaxies. As the majority of BCGs are at the exact same position as the COP of the host (by definition) their positions do not contribute to the shape estimate. We compared these shape estimates from the simulation with our result from the MC exercise by calculating the difference between the bias from the MC and the bias from estimating the shapes using satellite positions ($\Delta \ \mathrm{bias} = \mathrm{b}_{\mathrm{MC}} - \mathrm{b}_{\mathrm{sat}}$). We see that the MC reproduces the results from the simulations, with a $\Delta \ \mathrm{bias}$ of at most 0.13 for the 2.8 Gpc box, consistent with the values obtained for the smaller box. Halo axis ratios are on average biased low by 0.31 when 5 satellites are used to estimate the shape. If 50 satellite positions are used instead, this bias drops to 0.09. We see that the bias can be removed to a residual bias $\sim$ 0.1 at an $N_{\mathrm{sat}}$ value of 5, representing a decrease in the bias akin to almost using ten times more satellites to estimate the shape. This implies that around two-thirds of the bias can be explained by sampling.

In the middle and bottom panels of Fig.\ref{fig:sat_number-q_bias}, we also plot $\Delta \ \mathrm{bias}$ between the median bias from the MC exercise and the bias calculated from the simulations. The middle panel focuses on the different hydro-variations, and the bottom panel on different redshifts (of the 2.8 Gpc run). The redshift variation is also consistent with these results for the smaller 1 Gpc box (not shown). The error bars on the deviation from the MC are estimated using the delete one Jack-knife method \citep[for more details, see][]{Norberg2009}. We recalculated our bias variation with the number of satellites by omitting, in turn, a sub-volume of the simulation box. We used 27 sub-boxes, benefiting from the large-box size of the simulation. This was done for each feedback model and at each redshift for the $\mathrm{m}_9$ run.

We further compared the bias estimates when using the BCG or the mean of the satellite positions as the centre in the top and middle panels of Fig.\ref{fig:sat_number-q_bias}. Using the BCG as the centre increases the overall bias by around 0.05 and also deviates from the MC fit more at larger $N_{\mathrm{sat}}$ than when the centre is estimated simply by using the satellite distribution excluding the BCG. We also tested the MC with a fixed centre, which does not change the bias variation with $N_{\mathrm{sat}}$ significantly. Thus, we conclude that the BCG is a source of miscentering error, but that this effect is negligible for our purposes (as also noted by \citealt{Zhou2023}).

$\Delta \ \mathrm{bias}$ is around 0.1 at the low $N_{\mathrm{sat}}$ and consistent with 0 for the largest values. The hydrodynamic models are consistent with each other, showing that baryonic feedback does not change our results. Similarly, $\Delta \ \mathrm{bias}$ for different redshifts go from 0.1 to 0 with increasing $N_{\mathrm{sat}}$, and are consistent. Using the 2.8 Gpc box allows for better statistics especially for haloes with a large number of satellites. The variance on $\Delta \ \mathrm{bias}$ was estimated by sub-dividing the box into 8 sub-boxes, and taking the standard deviation of the resulting measurements. We find that $\Delta \ \mathrm{bias}$ is again around 0.1 at low $N_{\mathrm{sat}}$ and consistent with 0 for the largest values. Moreover, this trend remains consistent at higher redshifts. 

To first order, the shape bias is captured by the assumption of sampling noise, thus allowing us to remove the bias from the shapes of ensembles of haloes with as few as 5 satellites to less than 0.1 residual bias. This opens up the possibility of measuring the shapes of haloes (on average) which have lower masses than weak lensing methods are capable of measuring with sufficient signal to noise ratio. With our method, even though individual measurements are biased due to sampling noise, the ensemble estimate can be corrected for this bias. Alternatively, the simulations could also be used to correct the bias, as the bulk of this bias is understood.

\begin{figure}

    \centering
    \includegraphics[scale=0.4]{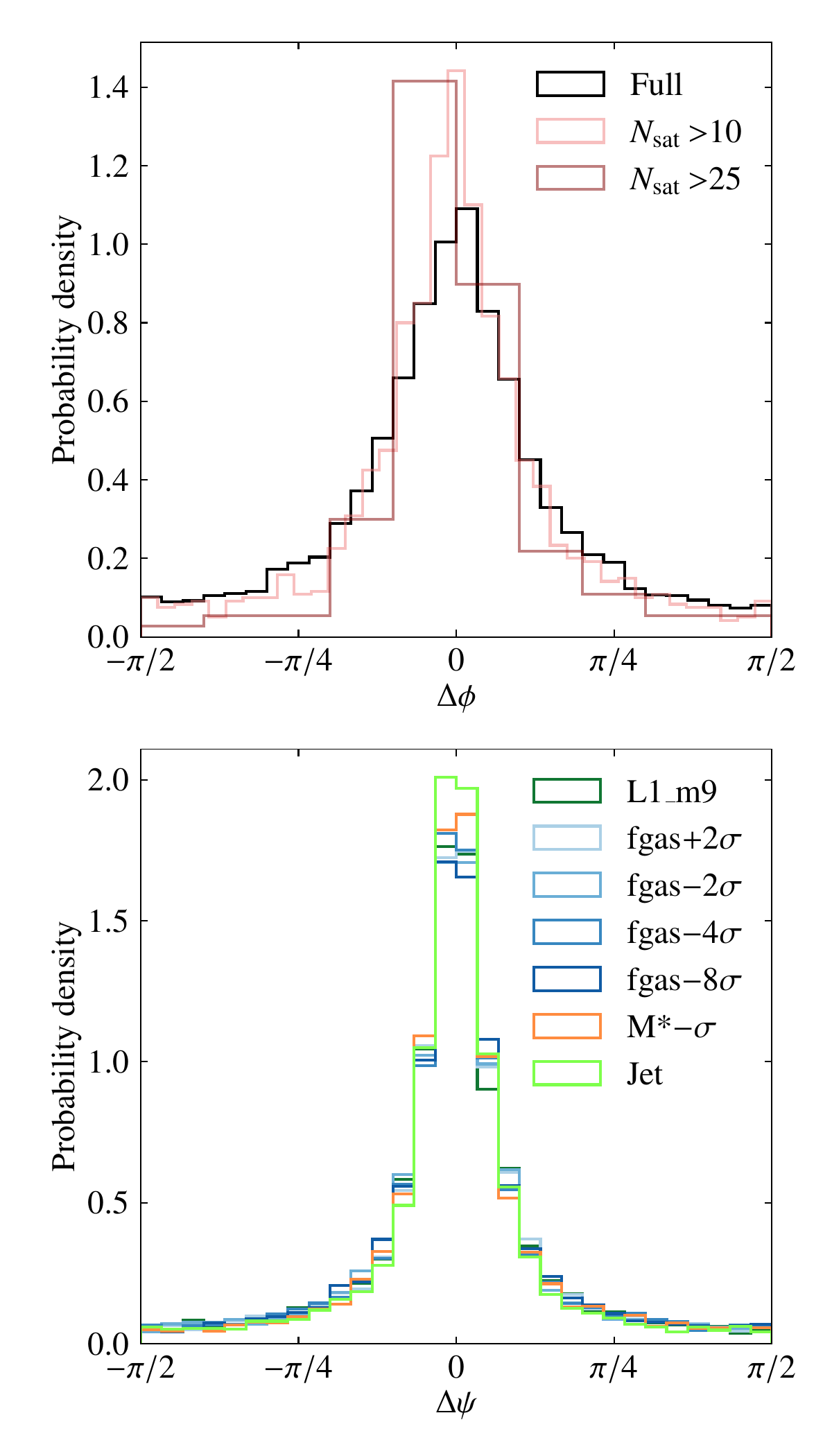}
    \caption{Top: Probability distribution of the misalignment angle in 2D between orientation angles calculated from the positions of satellite galaxies and the dark matter halo inertia tensor. The distribution after cuts are made on the minimum number of satellites are also shown. Bottom: Probability distribution of the misalignment angle in 2D between the BCG orientation angle and the dark matter halo inertia tensor for different feedback models.}
    \label{fig:misalignment_angles}

\end{figure}

\begin{figure*}

    \centering
    \includegraphics[scale=0.5]{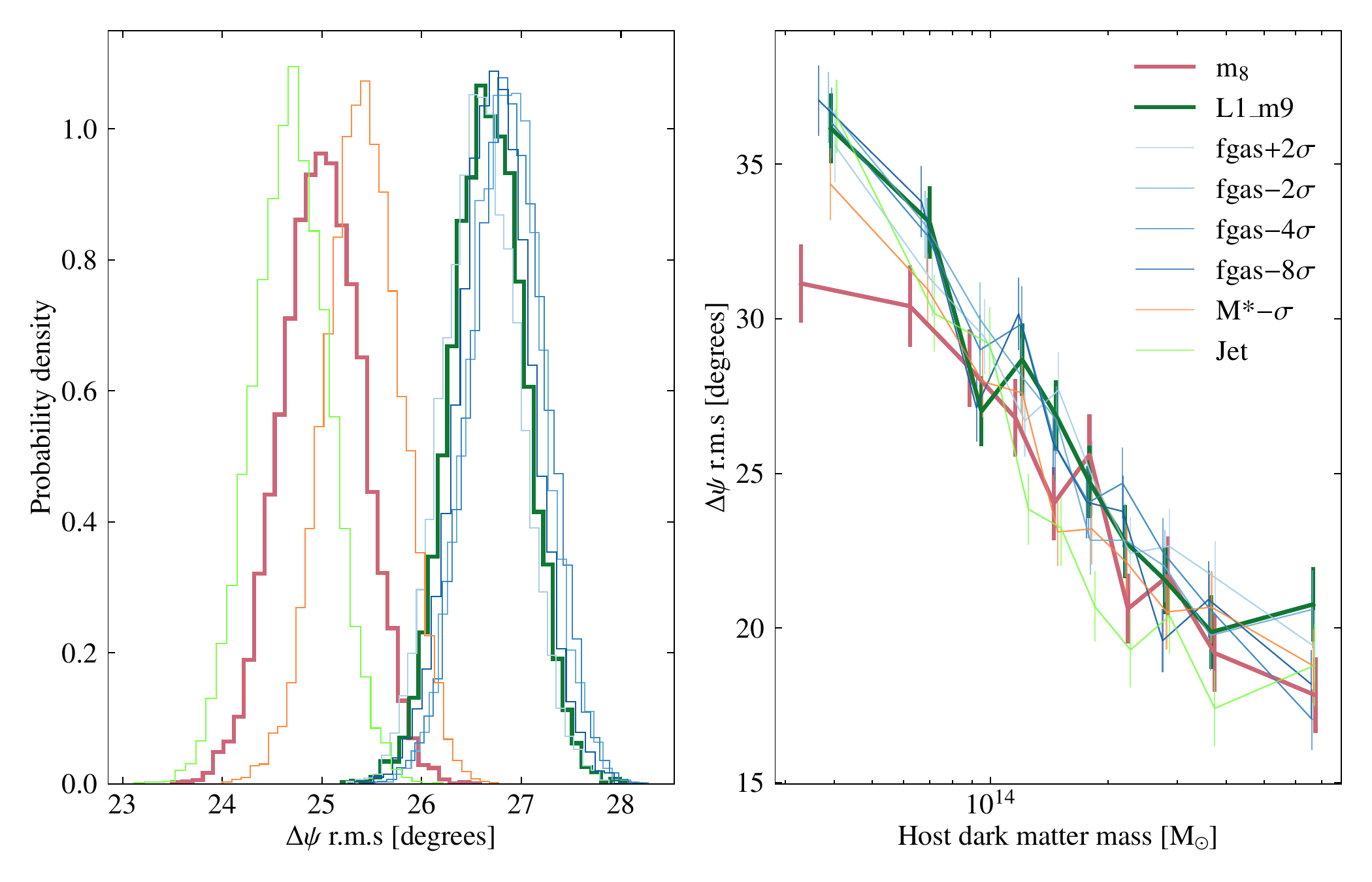}
    \caption{Left: Probability distribution of bootstrap realisations of the r.m.s of the projected misalignment angle between the BCG orientation angle and the dark matter halo inertia tensor. The different feedback variations are also shown. Right: Variation of the r.m.s of the projected misalignment angle between the BCG orientation angle and the dark matter halo inertia tensor with host halo dark matter (bound) mass.}
    \label{fig:del_psi_rms_distribution}

\end{figure*}

\begin{figure}

    \centering
    \includegraphics[width=0.5\textwidth]{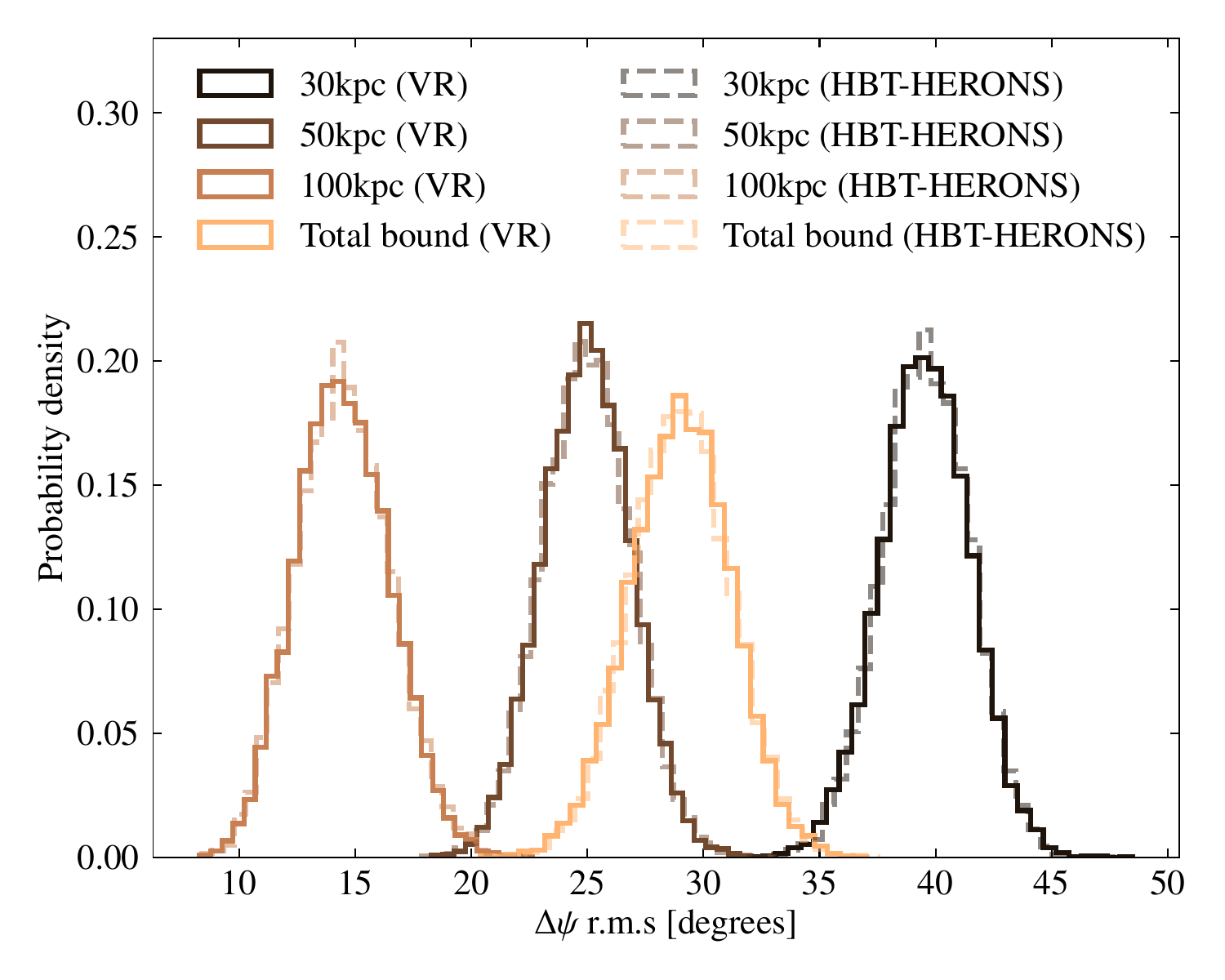}
    \caption{Probability distributions of bootstrapped realistions of the r.m.s of the projected misalignment angle between the BCG orientation angle and the dark matter halo at various projected apertures for the $\mathrm{m}_9$ fiducial run. Only haloes with at least 1000 stellar particles within 30 kpc are considered. The distributions calculated for both VR and HBT-HERONS are shown, with negligible differences. The apertures for which these values are calculated are 30, 50 and 100 kpc, as well as for the total halo.}
    \label{fig:bcg_host_misalignment_apertures}

\end{figure}
\subsection{Comparison with previous work}
\label{sect:comparison}

The bias from sampling noise on the estimates of halo shapes was also considered by \citet{Gonzalez2021}. Using clusters with mass greater than $10^{14}$ $\mathrm{M}_{\odot}$ from a hydrodynamical simulation, they showed that the positions of satellite galaxies trace the underlying dark matter distribution, even when there are very few members. They modelled the noise bias from having too few satellites by downsampling their haloes. They also found that discreteness noise tends to make haloes more elliptical. 
Our approach improves on their method by using a Monte Carlo setup, including realistic dark matter density profiles and shapes, in comparison with multiple resolution runs of hydrodynamical simulation to estimate how much of the bias comes from sampling noise. Additionally, our approach naturally incorporates the variation of the bias with the intrinsic shape of the halo, which is only captured in downsampling exercises if a large numbers of haloes with a broad range of realistic shapes are included.

A similar approach to this work was also implemented by \citet{Velliscig2015a} (see their Appendix A2). Triaxial NFWs with a constant concentration of $c_{\mathrm{NFW}} = 5$ were generated for constant sphericity $s = 0.6$ and triaxiality $t = 0.7$, and they found that sampling error leads to a bias of 0.02 for the sphericity at 300 particles. In this work, we have improved on this setup by incorporating variable shapes sampled from the distribution of shape parameters of the $\mathrm{m}_9$ run, in addition to a variable concentration that is related to the mass of the halo. The closest attempt to our work was carried out in Appendix A of \citet{Hoffmann2014}. Mock DM haloes with variable ellipticity and density profiles were generated, and it was found that around 1000 particles are required to estimate the shape to within 10 per cent, consistent with our findings. They also fitted a function to correct the derived axis ratios for sampling bias, but found that the improvements were only marginal for measuring dark matter halo shapes with subhaloes. However, their analysis was confined to four Milky Way-like host haloes with mass $\sim 10^{12} \ \mathrm{M}_{\odot}$. In this work, we have focused on more massive haloes and have included several thousand haloes in the analysis. 

We also compared with results from the FLAMINGO simulation, which has runs with identical initial conditions but different resolutions allowing us to measure resolution effects in the simulations. The large box size also allows us to have more than 100 times more cluster haloes than any previous analysis, with a comparison of the effects of different feedback prescriptions. The cross-validation with the simulation is invaluable and has not been attempted previously in the literature, allowing us to see that most of the bias is captured by modeling sampling noise. Previous works have been concerned with evaluating the level of bias on the shapes estimated when sampling bias is considered, in order to place a lower limit on the number of particles a halo must have in order for it to be included in the analysis. In this work, we show for the first time that this bias can be removed to less than 0.1 residual bias, thereby allowing us to push below the usual limit of $\approx 300 -1000$ particles for halo shape estimation. In addition to these improvements on previous works, the main contribution of this work is the realisation that the sampling bias from estimating dark matter halo shapes using member galaxy positions can be well-modelled with the same setup.

\section{Dark matter halo orientation angle}
\label{sect:misalignment}

Apart from the axis ratio, the orientation angle of haloes also has many applications. Stacking weak lensing shear maps can completely wash out the ellipticity signal, unless they are stacked along the orientation angle of the haloes \citep{Hoekstra2004, Evans2009}. IA studies also require an estimate of the orientation angle \citep{Zhou2023}. Similar to estimating the axis ratios of haloes, the orientation angle of the BCG can be adopted as the orientation angle of the host halo or it can be inferred from the positions of satellite galaxies. 

We investigated whether the satellite galaxies can be used to recover the orientation angle in the top panel of Fig.\ref{fig:misalignment_angles}. We show the distribution of the misalignment angle between orientation angles derived from the satellite shapes and those from the dark matter halo ($\Delta \phi$) for the $\mathrm{m}_9$ run. The black curve shows the full sample of 5627 haloes with at least 5 satellites with dark matter mass greater than $10^{12} \ \mathrm{M}_{\odot}$, while the other two histograms correspond to the misalignment calculated only for haloes with more than 10 and 25 member galaxies. The misalignment angle is centred around 0 with a non-zero plateau, and its distribution is non-Gaussian. Nonetheless, the root mean square (r.m.s) of the distribution is $33.19 \pm 0.40 \degree$. As we increase the cut on the number of satellite galaxies, the distribution becomes more strongly peaked at 0 with the non-zero plateau of the full distribution vanishing. We attribute this plateau to the random misalignment caused by sampling noise for shapes estimated using very few satellites.

The distribution of $\Delta \phi$ is consistent with previous results \citep[e.g][]{Agustsson2006, Shao2016} that show that satellite galaxies lie preferentially along the same axis as the dark matter halo. This can be explained by the fact that material streaming in from the filaments causes galaxies to lie on a preferred axis, i.e., in the direction of material infall. 

We also explore the misalignment between the orientation angle of the BCG with the host dark matter, $\Delta \psi$. Previous works disagree on the scatter of this misalignment. For example, \citet{Okumura2009} estimate a scatter of $35\degree$ for luminous red galaxies in the Sloan Digital Sky Survey. \citet{Herbonnet2022} prefer a lower scatter of $20\degree$ between the central galaxy orientation and the dark matter halo when using simulated clusters from the Three Hundred Project. \citet{Chisari2017} find a range of possible values depending on the mass of the host halo in the Horizon-AGN cosmological hydrodynamical simulation. When restricted to the typical halo mass of luminous red galaxies, they obtain a scatter of $48\degree$. Generally speaking, these studies find sensitivity of the misalignment angle scatter with host halo mass, which could be driving a disagreement between their results. Moreover, the sample sizes were much smaller than ours, and thus possibly affected by cosmic variance.

In the bottom panel of Fig.\ref{fig:misalignment_angles}, the distribution of $\Delta \psi$ is shown for the $\mathrm{m}_9$ run, for all baryonic feedback models. We find that the overall scatter of this distribution is lower than for $\Delta \phi$, with the fiducial feedback model for the highest resolution run having an r.m.s of $25.01 \pm 0.40 \degree$. We show the distributions of bootstrapped realisations of the r.m.s of $\Delta \psi$ in the left panel of Fig.\ref{fig:del_psi_rms_distribution}. We can see that the Jet model has the lowest scatter, but overall the feedback model has only a small impact on the value of $\Delta \psi$. The comparison with the $\mathrm{m}_8$ run in the right panel of Fig.\ref{fig:del_psi_rms_distribution} reveals that this measurement also seems to be affected by resolution effects, especially at lower halo masses. Although the impact of resolution on the alignment angle is significant, we note that it only shifts by a few degrees. In the alignment correlation function, this should translate into a small amplitude change compared to the $\mathrm{m}_9$ case. We will explore this phenomenon more deeply in future work.

With the large number of haloes we have in this analysis, we can also explore the mass dependence of $\Delta \psi$. This is shown in the right panel of Fig.\ref{fig:del_psi_rms_distribution} for the various feedback models. We see a decrease in the r.m.s of $\Delta \psi$ from $35 \degree$ at the low mass end down to about $20 \degree$ for the most massive haloes. This allows us to conclude that the mass dependence as well as the poorer statistics of some previous studies might explain the discrepancy of the scatter values they obtained. In fact, both \citet{Tenneti2014} and \citet{Chisari2017} find a decrease in the misalignment angle with increasing halo mass, similar to our results. This is further corroborated by the findings of \citet{Velliscig2015a} who find a median misalignment angle of $45-50\degree$ for haloes with mass in the range $11 \le \log_{10}(M_{200}/[h^{-1}\mathrm{M}_{\odot}]) \le 15$, which drops to $25-30\degree$ when restricting the analysis to the most massive haloes ($13 \le \log_{10}(M_{200}/[h^{-1}\mathrm{M}_{\odot}]) \le 15$), which is in very good agreement with our results.

In addition to the mass dependence of $\Delta \psi$, we also studied its radial dependence. In Fig.\ref{fig:bcg_host_misalignment_apertures}, we show the misalignment angle between the BCG and the dark matter distribution of the host as a function of radius. To ensure that the shapes included in the analysis are well resolved, we selected haloes that have a minimum number of particles within 30 kpc. The choice of this cut was motivated by the bias on the scatter of the misalignment angle as a function of particle number from our MC, which reaches 0 only for $N > 10^3$ particles, shown in Fig.\ref{fig:position_angle_bias}. As there were no haloes satisfying a cut of $10^{4}$ particles within 30 kpc, we chose $10^{3}$ as the particle cut. Although this might seem very conservative, our MC only captures the bias on the orientation angle from sampling noise, and the comparison with the different resolution runs of the simulation revealed an additional scatter from the removal of lower mass substructures from a halo in a higher resolution run compared to its lower resolution version. This is explored in more detail in Appendix \ref{sect:appendixB}. We also compared the distributions calculated for haloes found with VR and HBT-HERONS. We found very little difference between the two halo finders, which we can attribute to the conservative particle cut. 

Fig.\ref{fig:bcg_host_misalignment_apertures} shows that the misalignment angle is largest within 30 kpc. However, this is about 3 softening lengths, and thus we conclude that this measurement is unreliable. At larger apertures, well outside the spatial resolution limit of the simulation, the misalignment angle is lower, with the 100 kpc having the lowest misalignment. This is because haloes are more spherical closer to the centre, which results in larger misalignment angles at smaller apertures. 100 kpc seems to be the scale at which the axis ratios of the stellar component of the BCG and the dark matter distribution of the host are elliptical enough to lower the misalignment angle. Since the $\Delta \phi$ has a larger scatter than $\Delta \psi$, the orientation angle of the dark matter halo is better estimated using the BCG orientations. Moreover, the orientation of the BCG also has a strong radial dependence, and thus the aperture within which the orientation angle is estimated will influence the results.

The radial dependence of the misalignment angle was also explored by \citet{Velliscig2015a}, who found that the total matter has a very low misalignment with the stellar component. This was motivated by the fact that weak lensing is sensitive to the total matter. However, the total matter is of course correlated with the stellar component, thus motivating our comparison of the stellar component to the dark matter alone. This also means that we cannot make a one-to-one comparison with their work.    

\section{Discussion \& Conclusions}
\label{sect:disc_conc}

Halo shapes are valuable probes of both cosmology and baryonic feedback, but are challenging to infer observationally. Weak lensing is sensitive to dark matter halo properties, and provides a method to infer their shapes using galaxy survey data. However, the signal is typically too low for individual halo shapes to be measured, and we must rely on results from stacking \citep[see][for a detailed review on the topic]{Hoekstra2013}. Directly linking the positions of satellite galaxies to the shape of their host is thus an important step in the reliable estimation of halo shapes.

However, this method suffers from a lack of a large number of satellite galaxies per host halo, thereby causing the derived axis ratios to be biased. Using our MC setup, we saw that projected halo shapes appear more elliptical than they really are due to sampling noise. We have shown that this error can be estimated and thus corrected for, thereby allowing us to estimate, on average, the shapes of dark matter haloes that have even as few as 5 galaxies residing within them (Fig.\ref{fig:sat_number-q_bias}). This enables the measurement of shapes of haloes with lower masses than weak lensing methods can probe. 

We made use of the large statistical power of the FLAMINGO runs, as well as the ability to understand resolution effects with the multiple mass resolutions implemented. This enabled a verification of the axis ratio bias correction for runs at different feedback variations and redshifts. We saw that including the BCG as the centre of the ellipse actually increases the overall bias. Determining halo shapes with the satellite positions while leaving the centre free results in a lower bias, as well as a better correction using our bias estimate. 

Apart from the axis ratio, the orientation angle of the projected halo ellipse is also of interest. We found that the misalignment angle between the orientation angles of the satellite positions and the host halo has a larger scatter than the misalignment between the halo and the BCG orientation angles (see Fig.\ref{fig:misalignment_angles}). Moreover, this orientation angle is affected by having too few satellites, which serves to randomise its direction. We also reconciled previous measurements of the scatter of the misalignment between the BCG and host halo orientations using the large number of haloes available, which revealed a strong correlation between this quantity and the mass of the host halo (Fig.\ref{fig:del_psi_rms_distribution}), in good agreement with the trends evidenced in smaller hydrodynamical simulations \citet{Tenneti2014, Velliscig2015a, Chisari2017}, where the stellar component was strongly aligned with the dark matter one. 

For future surveys, it is thus probably more prudent to estimate the shapes of haloes using the positions of satellites, with the halo centre also estimated using the satellite positions, while reserving the BCG to estimate the orientation angle of the ellipse. However, in this study, we have not incorporated certain observational effects that could influence our choice of shape and angle tracers. For example, any analysis using the positions of satellites to estimate halo shape requires that the membership probability of each galaxy needs to be above a certain threshold to be considered a tracer of that halo. This arises from the fact that photometric redshifts are used to determine whether a galaxy is a member of a specific cluster or not, together with codes like redMaPPer, which is a cluster finding algorithm \citep{Rykoff2014} that provides membership probabilities for each galaxy. When estimating shapes of haloes with member galaxies, a minimum cut is placed on this probability with considerations towards purity of the sample and number of satellites per cluster \citep{vanUitert2017b}. Although this type of analysis could benefit from our bias correction as halo shapes with even very strict cuts on membership probability (which allow only about 5 tracers per halo) could be corrected, we have not explored the possible effects of line-of-sight contamination in this work. In addition, although the halo orientation angle is determined better using the BCG as opposed to the satellite galaxies, identifying the BCG of a cluster is a challenging task \citep{Lauer2014}. Misidentifying the halo centre can lead to centering errors for the determination of the orientation angle, which is problematic for modeling of the IA signal.

In order to determine to what extent these observational challenges affect the halo shape measurements, shapes derived from weak lensing and by using the intra-cluster light \citep{Gonzalez2021}, can be compared to those derived from satellite positions, using data from the same survey. One could also make mock observations of haloes from FLAMINGO and compare with the shapes derived from the various methods. These are beyond the scope of this work and we leave it to a possible future analysis. We also explored the possibility of correcting the halo orientation angles in an appendix. We concluded that increasing the resolution of a simulation increases the number of lower mass substructures that are resolved, which is an additional scatter that is difficult to capture.

In conclusion, correctly accounting for sampling noise allows the recovery of unbiased estimates of axis ratios of the host dark matter haloes using satellite galaxies as tracers. This paves the way for accurate shape measurements of ensembles of dark matter haloes in future surveys. Moreover, this method can also be used to correct shapes of haloes in simulations statistically, thereby allowing the lower limit on the halo particle number to be reduced. This can be particularly useful in the case of IA studies, as one could benefit from the large number of haloes that can now be included in the analysis, which were previously excluded due to the concern of biased shape estimates. In addition, this also informs future IA studies with cosmological simulations, as one could prioritise larger box sizes over mass resolution, instead relying on correcting biased shapes of haloes with very few particles. In the future, this methodology could also be extended to the correction of galaxy shapes in simulations for galactic-scale IA studies.

\begin{acknowledgements}
AH thanks Jeger Broxterman, Bianca Sersante, Divya Rana and Will McDonald for valuable discussions. AH acknowledges support by NWO through the Dark Universe Science Collaboration (OCENW.XL21.XL21.025). 

NEC acknowledges support from the project ``A rising tide: Galaxy intrinsic alignments as a new probe of cosmology and galaxy evolution'' (with project number VI.Vidi.203.011) of the Talent programme Vidi which is (partly) financed by the Dutch Research Council (NWO). 

HH acknowledges funding from the European Research Council (ERC) under the European Union's Horizon 2020 research and innovation program (Grant agreement No. 101053992).

This work used the DiRAC@Durham facility managed by the Institute for Computational Cosmology on behalf of the STFC DiRAC HPC Facility (www.dirac.ac.uk). The equipment was funded by BEIS capital funding via STFC capital grants ST/K00042X/1, ST/P002293/1, ST/R002371/1 and ST/S002502/1, Durham University and STFC operations grant ST/R000832/1. DiRAC is part of the National e-Infrastructure. 

This work was done using PYTHON (\url{http://www.python.org}), employing the packages NUMPY \citep{Harris2020}, MATPLOTLIB \citep{Hunter2007}, SCIPY \citep{Scipy2020} and CCL \citep{Chisari2019}. The NFW profile is sampled using the package detailed by \citet{Robotham2018}.

\end{acknowledgements}

%
%
\bibliographystyle{aa} 
\bibliography{main} 

\begin{thebibliography}{74}
\expandafter\ifx\csname natexlab\endcsname\relax\def\natexlab#1{#1}\fi

\bibitem[{{Agustsson} \& {Brainerd}(2006)}]{Agustsson2006}
{Agustsson}, I. \& {Brainerd}, T.~G. 2006, \apjl, 644, L25

\bibitem[{{Allgood} {et~al.}(2006){Allgood}, {Flores}, {Primack}, {Kravtsov},
  {Wechsler}, {Faltenbacher}, \& {Bullock}}]{Allgood2006}
{Allgood}, B., {Flores}, R.~A., {Primack}, J.~R., {et~al.} 2006, \mnras, 367,
  1781

\bibitem[{{Bett}(2012)}]{Bett2012}
{Bett}, P. 2012, \mnras, 420, 3303

\bibitem[{{Booth} \& {Schaye}(2009)}]{Booth2009}
{Booth}, C.~M. \& {Schaye}, J. 2009, \mnras, 398, 53

\bibitem[{{Borrow} {et~al.}(2022){Borrow}, {Schaller}, {Bower}, \&
  {Schaye}}]{Borrow2022}
{Borrow}, J., {Schaller}, M., {Bower}, R.~G., \& {Schaye}, J. 2022, \mnras,
  511, 2367

\bibitem[{{Bryan} {et~al.}(2013){Bryan}, {Kay}, {Duffy}, {Schaye}, {Dalla
  Vecchia}, \& {Booth}}]{Bryan2013}
{Bryan}, S.~E., {Kay}, S.~T., {Duffy}, A.~R., {et~al.} 2013, \mnras, 429, 3316

\bibitem[{{Catelan} {et~al.}(2001){Catelan}, {Kamionkowski}, \&
  {Blandford}}]{Catelan2001}
{Catelan}, P., {Kamionkowski}, M., \& {Blandford}, R.~D. 2001, \mnras, 320, L7

\bibitem[{{Chaikin} {et~al.}(2023){Chaikin}, {Schaye}, {Schaller},
  {Ben{\'\i}tez-Llambay}, {Nobels}, \& {Ploeckinger}}]{Chaikin2023}
{Chaikin}, E., {Schaye}, J., {Schaller}, M., {et~al.} 2023, \mnras, 523, 3709

\bibitem[{{Chisari} {et~al.}(2015){Chisari}, {Codis}, {Laigle}, {Dubois},
  {Pichon}, {Devriendt}, {Slyz}, {Miller}, {Gavazzi}, \&
  {Benabed}}]{Chisari2015}
{Chisari}, N., {Codis}, S., {Laigle}, C., {et~al.} 2015, \mnras, 454, 2736

\bibitem[{{Chisari} {et~al.}(2016){Chisari}, {Laigle}, {Codis}, {Dubois},
  {Devriendt}, {Miller}, {Benabed}, {Slyz}, {Gavazzi}, \&
  {Pichon}}]{Chisari2016}
{Chisari}, N., {Laigle}, C., {Codis}, S., {et~al.} 2016, \mnras, 461, 2702

\bibitem[{{Chisari} {et~al.}(2019){Chisari}, {Alonso}, {Krause}, {Leonard},
  {Bull}, {Neveu}, {Villarreal}, {Singh}, {McClintock}, {Ellison}, {Du},
  {Zuntz}, {Mead}, {Joudaki}, {Lorenz}, {Tr{\"o}ster}, {Sanchez}, {Lanusse},
  {Ishak}, {Hlozek}, {Blazek}, {Campagne}, {Almoubayyed}, {Eifler}, {Kirby},
  {Kirkby}, {Plaszczynski}, {Slosar}, {Vrastil}, {Wagoner}, \& {LSST Dark
  Energy Science Collaboration}}]{Chisari2019}
{Chisari}, N.~E., {Alonso}, D., {Krause}, E., {et~al.} 2019, \apjs, 242, 2

\bibitem[{{Chisari} {et~al.}(2017){Chisari}, {Koukoufilippas}, {Jindal},
  {Peirani}, {Beckmann}, {Codis}, {Devriendt}, {Miller}, {Dubois}, {Laigle},
  {Slyz}, \& {Pichon}}]{Chisari2017}
{Chisari}, N.~E., {Koukoufilippas}, N., {Jindal}, A., {et~al.} 2017, \mnras,
  472, 1163

\bibitem[{{Dalla Vecchia} \& {Schaye}(2008)}]{DallaVecchia2008}
{Dalla Vecchia}, C. \& {Schaye}, J. 2008, \mnras, 387, 1431

\bibitem[{{Despali} {et~al.}(2014){Despali}, {Giocoli}, \&
  {Tormen}}]{Despali2014}
{Despali}, G., {Giocoli}, C., \& {Tormen}, G. 2014, \mnras, 443, 3208

\bibitem[{{Dubinski} \& {Carlberg}(1991)}]{Dubinski1991}
{Dubinski}, J. \& {Carlberg}, R.~G. 1991, \apj, 378, 496

\bibitem[{{Elahi} {et~al.}(2019){Elahi}, {Ca{\~n}as}, {Poulton}, {Tobar},
  {Willis}, {Lagos}, {Power}, \& {Robotham}}]{Elahi2019}
{Elahi}, P.~J., {Ca{\~n}as}, R., {Poulton}, R. J.~J., {et~al.} 2019, \pasa, 36,
  e021

\bibitem[{{Elbers} {et~al.}(2021){Elbers}, {Frenk}, {Jenkins}, {Li}, \&
  {Pascoli}}]{Elbers2021}
{Elbers}, W., {Frenk}, C.~S., {Jenkins}, A., {Li}, B., \& {Pascoli}, S. 2021,
  \mnras, 507, 2614

\bibitem[{{Evans} \& {Bridle}(2009)}]{Evans2009}
{Evans}, A. K.~D. \& {Bridle}, S. 2009, \apj, 695, 1446

\bibitem[{{Fortuna} {et~al.}(2021){Fortuna}, {Hoekstra}, {Joachimi},
  {Johnston}, {Chisari}, {Georgiou}, \& {Mahony}}]{Fortuna2021}
{Fortuna}, M.~C., {Hoekstra}, H., {Joachimi}, B., {et~al.} 2021, \mnras, 501,
  2983

\bibitem[{{Georgiou} {et~al.}(2021){Georgiou}, {Hoekstra}, {Kuijken},
  {Bilicki}, {Dvornik}, {Erben}, {Giblin}, {Heymans}, {Hildebrandt}, {de Jong},
  {Kannawadi}, {Schneider}, {Schrabback}, {Shan}, \& {Wright}}]{Georgiou2021}
{Georgiou}, C., {Hoekstra}, H., {Kuijken}, K., {et~al.} 2021, \aap, 647, A185

\bibitem[{{Gonzalez} {et~al.}(2021){Gonzalez}, {Ragone-Figueroa}, {Donzelli},
  {Makler}, {Garc{\'\i}a Lambas}, \& {Granato}}]{Gonzalez2021}
{Gonzalez}, E.~J., {Ragone-Figueroa}, C., {Donzelli}, C.~J., {et~al.} 2021,
  \mnras, 508, 1280

\bibitem[{{Gonzalez} {et~al.}(2024){Gonzalez}, {Rodr{\'\i}guez-Medrano},
  {Pereyra}, \& {Garc{\'\i}a Lambas}}]{Gonzalez2024}
{Gonzalez}, E.~J., {Rodr{\'\i}guez-Medrano}, A., {Pereyra}, L., \& {Garc{\'\i}a
  Lambas}, D. 2024, \mnras, 528, 3075

\bibitem[{{Han} {et~al.}(2018){Han}, {Cole}, {Frenk}, {Benitez-Llambay}, \&
  {Helly}}]{Han2018}
{Han}, J., {Cole}, S., {Frenk}, C.~S., {Benitez-Llambay}, A., \& {Helly}, J.
  2018, \mnras, 474, 604

\bibitem[{{Harris} {et~al.}(2020){Harris}, {Millman}, {van der Walt},
  {Gommers}, {Virtanen}, {Cournapeau}, {Wieser}, {Taylor}, {Berg}, {Smith},
  {Kern}, {Picus}, {Hoyer}, {van Kerkwijk}, {Brett}, {Haldane}, {del R{\'\i}o},
  {Wiebe}, {Peterson}, {G{\'e}rard-Marchant}, {Sheppard}, {Reddy}, {Weckesser},
  {Abbasi}, {Gohlke}, \& {Oliphant}}]{Harris2020}
{Harris}, C.~R., {Millman}, K.~J., {van der Walt}, S.~J., {et~al.} 2020, \nat,
  585, 357

\bibitem[{{Herbonnet} {et~al.}(2022){Herbonnet}, {Crawford}, {Avestruz},
  {Rasia}, {Giocoli}, {Meneghetti}, {von der Linden}, {Cui}, \&
  {Yepes}}]{Herbonnet2022}
{Herbonnet}, R., {Crawford}, A., {Avestruz}, C., {et~al.} 2022, \mnras, 513,
  2178

\bibitem[{{Hirata} {et~al.}(2007){Hirata}, {Mandelbaum}, {Ishak}, {Seljak},
  {Nichol}, {Pimbblet}, {Ross}, \& {Wake}}]{Hirata2007}
{Hirata}, C.~M., {Mandelbaum}, R., {Ishak}, M., {et~al.} 2007, \mnras, 381,
  1197

\bibitem[{{Hirata} \& {Seljak}(2004)}]{Hirata2004}
{Hirata}, C.~M. \& {Seljak}, U. 2004, \prd, 70, 063526

\bibitem[{{Ho} {et~al.}(2006){Ho}, {Bahcall}, \& {Bode}}]{Ho2006}
{Ho}, S., {Bahcall}, N., \& {Bode}, P. 2006, \apj, 647, 8

\bibitem[{{Hoekstra}(2013)}]{Hoekstra2013}
{Hoekstra}, H. 2013, arXiv e-prints, arXiv:1312.5981

\bibitem[{{Hoekstra} {et~al.}(1998){Hoekstra}, {Franx}, {Kuijken}, \&
  {Squires}}]{Hoekstra1998}
{Hoekstra}, H., {Franx}, M., {Kuijken}, K., \& {Squires}, G. 1998, \apj, 504,
  636

\bibitem[{{Hoekstra} {et~al.}(2004){Hoekstra}, {Yee}, \&
  {Gladders}}]{Hoekstra2004}
{Hoekstra}, H., {Yee}, H.~K.~C., \& {Gladders}, M.~D. 2004, \apj, 606, 67

\bibitem[{{Hoffmann} {et~al.}(2014){Hoffmann}, {Planelles}, {Gazta{\~n}aga},
  {Knebe}, {Pearce}, {Lux}, {Onions}, {Muldrew}, {Elahi}, {Behroozi},
  {Ascasibar}, {Han}, {Maciejewski}, {Merchan}, {Neyrinck}, {Ruiz}, \&
  {Sgro}}]{Hoffmann2014}
{Hoffmann}, K., {Planelles}, S., {Gazta{\~n}aga}, E., {et~al.} 2014, \mnras,
  442, 1197

\bibitem[{{Hunter}(2007)}]{Hunter2007}
{Hunter}, J.~D. 2007, Computing in Science and Engineering, 9, 90

\bibitem[{{Hu{\v{s}}ko} {et~al.}(2022){Hu{\v{s}}ko}, {Lacey}, {Schaye},
  {Schaller}, \& {Nobels}}]{Husko2022}
{Hu{\v{s}}ko}, F., {Lacey}, C.~G., {Schaye}, J., {Schaller}, M., \& {Nobels},
  F. S.~J. 2022, \mnras, 516, 3750

\bibitem[{{Jeeson-Daniel} {et~al.}(2011){Jeeson-Daniel}, {Dalla Vecchia},
  {Haas}, \& {Schaye}}]{Jeeson-Daniel2011}
{Jeeson-Daniel}, A., {Dalla Vecchia}, C., {Haas}, M.~R., \& {Schaye}, J. 2011,
  \mnras, 415, L69

\bibitem[{{Jing} \& {Suto}(2002)}]{Jing2002}
{Jing}, Y.~P. \& {Suto}, Y. 2002, \apj, 574, 538

\bibitem[{{Joachimi} {et~al.}(2015){Joachimi}, {Cacciato}, {Kitching},
  {Leonard}, {Mandelbaum}, {Sch{\"a}fer}, {Sif{\'o}n}, {Hoekstra}, {Kiessling},
  {Kirk}, \& {Rassat}}]{Joachimi2015}
{Joachimi}, B., {Cacciato}, M., {Kitching}, T.~D., {et~al.} 2015, \ssr, 193, 1

\bibitem[{{Kugel} {et~al.}(2023){Kugel}, {Schaye}, {Schaller}, {Helly},
  {Braspenning}, {Elbers}, {Frenk}, {McCarthy}, {Kwan}, {Salcido}, {van
  Daalen}, {Vandenbroucke}, {Bah{\'e}}, {Borrow}, {Chaikin}, {Hu{\v{s}}ko},
  {Jenkins}, {Lacey}, {Nobels}, \& {Vernon}}]{Kugel2023}
{Kugel}, R., {Schaye}, J., {Schaller}, M., {et~al.} 2023, \mnras, 526, 6103

\bibitem[{{Lau} {et~al.}(2021){Lau}, {Hearin}, {Nagai}, \&
  {Cappelluti}}]{Lau2021}
{Lau}, E.~T., {Hearin}, A.~P., {Nagai}, D., \& {Cappelluti}, N. 2021, \mnras,
  500, 1029

\bibitem[{{Lauer} {et~al.}(2014){Lauer}, {Postman}, {Strauss}, {Graves}, \&
  {Chisari}}]{Lauer2014}
{Lauer}, T.~R., {Postman}, M., {Strauss}, M.~A., {Graves}, G.~J., \& {Chisari},
  N.~E. 2014, \apj, 797, 82

\bibitem[{{Mandelbaum} {et~al.}(2006){Mandelbaum}, {Hirata}, {Broderick},
  {Seljak}, \& {Brinkmann}}]{Mandelbaum2006}
{Mandelbaum}, R., {Hirata}, C.~M., {Broderick}, T., {Seljak}, U., \&
  {Brinkmann}, J. 2006, \mnras, 370, 1008

\bibitem[{{Melchior} \& {Viola}(2012)}]{Melchior2012}
{Melchior}, P. \& {Viola}, M. 2012, \mnras, 424, 2757

\bibitem[{Moreno {et~al.}(2025)Moreno, Helly, McGibbon, Schaye, Schaller, Han,
  \& Kugel}]{Moreno2025}
Moreno, V. J.~F., Helly, J., McGibbon, R., {et~al.} 2025, Assessing subhalo
  finders in cosmological hydrodynamical simulations

\bibitem[{{Navarro} {et~al.}(1996){Navarro}, {Frenk}, \& {White}}]{Navarro1996}
{Navarro}, J.~F., {Frenk}, C.~S., \& {White}, S. D.~M. 1996, \apj, 462, 563

\bibitem[{{Navarro} {et~al.}(1997){Navarro}, {Frenk}, \& {White}}]{Navarro1997}
{Navarro}, J.~F., {Frenk}, C.~S., \& {White}, S. D.~M. 1997, \apj, 490, 493

\bibitem[{{Norberg} {et~al.}(2009){Norberg}, {Baugh}, {Gazta{\~n}aga}, \&
  {Croton}}]{Norberg2009}
{Norberg}, P., {Baugh}, C.~M., {Gazta{\~n}aga}, E., \& {Croton}, D.~J. 2009,
  \mnras, 396, 19

\bibitem[{{Oguri} {et~al.}(2010){Oguri}, {Takada}, {Okabe}, \&
  {Smith}}]{Oguri2010}
{Oguri}, M., {Takada}, M., {Okabe}, N., \& {Smith}, G.~P. 2010, \mnras, 405,
  2215

\bibitem[{{Okumura} {et~al.}(2009){Okumura}, {Jing}, \& {Li}}]{Okumura2009}
{Okumura}, T., {Jing}, Y.~P., \& {Li}, C. 2009, \apj, 694, 214

\bibitem[{{Peter} {et~al.}(2013){Peter}, {Rocha}, {Bullock}, \&
  {Kaplinghat}}]{Peter2013}
{Peter}, A. H.~G., {Rocha}, M., {Bullock}, J.~S., \& {Kaplinghat}, M. 2013,
  \mnras, 430, 105

\bibitem[{{Ploeckinger} \& {Schaye}(2020)}]{Ploeckinger2020}
{Ploeckinger}, S. \& {Schaye}, J. 2020, \mnras, 497, 4857

\bibitem[{{Ragone-Figueroa} {et~al.}(2010){Ragone-Figueroa}, {Plionis},
  {Merch{\'a}n}, {Gottl{\"o}ber}, \& {Yepes}}]{Ragone-Figueroa2010}
{Ragone-Figueroa}, C., {Plionis}, M., {Merch{\'a}n}, M., {Gottl{\"o}ber}, S.,
  \& {Yepes}, G. 2010, \mnras, 407, 581

\bibitem[{{Robertson} {et~al.}(2019){Robertson}, {Harvey}, {Massey}, {Eke},
  {McCarthy}, {Jauzac}, {Li}, \& {Schaye}}]{Robertson2019}
{Robertson}, A., {Harvey}, D., {Massey}, R., {et~al.} 2019, \mnras, 488, 3646

\bibitem[{{Robotham} \& {Howlett}(2018)}]{Robotham2018}
{Robotham}, A.~S.~G. \& {Howlett}, C. 2018, Research Notes of the American
  Astronomical Society, 2, 55

\bibitem[{{Rykoff} {et~al.}(2014){Rykoff}, {Rozo}, {Busha}, {Cunha},
  {Finoguenov}, {Evrard}, {Hao}, {Koester}, {Leauthaud}, {Nord}, {Pierre},
  {Reddick}, {Sadibekova}, {Sheldon}, \& {Wechsler}}]{Rykoff2014}
{Rykoff}, E.~S., {Rozo}, E., {Busha}, M.~T., {et~al.} 2014, \apj, 785, 104

\bibitem[{{Schaller} {et~al.}(2024){Schaller}, {Borrow}, {Draper}, {Ivkovic},
  {McAlpine}, {Vandenbroucke}, {Bah{\'e}}, {Chaikin}, {Chalk}, {Chan},
  {Correa}, {van Daalen}, {Elbers}, {Gonnet}, {Hausammann}, {Helly},
  {Hu{\v{s}}ko}, {Kegerreis}, {Nobels}, {Ploeckinger}, {Revaz}, {Roper},
  {Ruiz-Bonilla}, {Sandnes}, {Uyttenhove}, {Willis}, \& {Xiang}}]{Schaller2024}
{Schaller}, M., {Borrow}, J., {Draper}, P.~W., {et~al.} 2024, \mnras
  [\eprint[arXiv]{2305.13380}]

\bibitem[{{Schaye} {et~al.}(2015){Schaye}, {Crain}, {Bower}, {Furlong},
  {Schaller}, {Theuns}, {Dalla Vecchia}, {Frenk}, {McCarthy}, {Helly},
  {Jenkins}, {Rosas-Guevara}, {White}, {Baes}, {Booth}, {Camps}, {Navarro},
  {Qu}, {Rahmati}, {Sawala}, {Thomas}, \& {Trayford}}]{Schaye2015}
{Schaye}, J., {Crain}, R.~A., {Bower}, R.~G., {et~al.} 2015, \mnras, 446, 521

\bibitem[{{Schaye} \& {Dalla Vecchia}(2008)}]{Schaye2008}
{Schaye}, J. \& {Dalla Vecchia}, C. 2008, \mnras, 383, 1210

\bibitem[{{Schaye} {et~al.}(2023){Schaye}, {Kugel}, {Schaller}, {Helly},
  {Braspenning}, {Elbers}, {McCarthy}, {van Daalen}, {Vandenbroucke}, {Frenk},
  {Kwan}, {Salcido}, {Bah{\'e}}, {Borrow}, {Chaikin}, {Hahn}, {Hu{\v{s}}ko},
  {Jenkins}, {Lacey}, \& {Nobels}}]{Schaye2023}
{Schaye}, J., {Kugel}, R., {Schaller}, M., {et~al.} 2023, \mnras, 526, 4978

\bibitem[{{Schneider} \& {Bridle}(2010)}]{Schneider2010}
{Schneider}, M.~D. \& {Bridle}, S. 2010, \mnras, 402, 2127

\bibitem[{{Schneider} {et~al.}(2012){Schneider}, {Frenk}, \&
  {Cole}}]{Schneider2012}
{Schneider}, M.~D., {Frenk}, C.~S., \& {Cole}, S. 2012, \jcap, 2012, 030

\bibitem[{{Schrabback} {et~al.}(2021){Schrabback}, {Hoekstra}, {Van Waerbeke},
  {van Uitert}, {Georgiou}, {Asgari}, {C{\^o}t{\'e}}, {Cuillandre}, {Erben},
  {Ferrarese}, {Gwyn}, {Heymans}, {Hildebrandt}, {Kannawadi}, {Kuijken},
  {Leauthaud}, {Makler}, {Mei}, {Miller}, {Raichoor}, {Schneider}, \&
  {Wright}}]{Schrabback2021}
{Schrabback}, T., {Hoekstra}, H., {Van Waerbeke}, L., {et~al.} 2021, \aap, 646,
  A73

\bibitem[{{Sereno} {et~al.}(2018){Sereno}, {Umetsu}, {Ettori}, {Sayers},
  {Chiu}, {Meneghetti}, {Vega-Ferrero}, \& {Zitrin}}]{Sereno2018}
{Sereno}, M., {Umetsu}, K., {Ettori}, S., {et~al.} 2018, \apjl, 860, L4

\bibitem[{{Shao} {et~al.}(2016){Shao}, {Cautun}, {Frenk}, {Gao}, {Crain},
  {Schaller}, {Schaye}, \& {Theuns}}]{Shao2016}
{Shao}, S., {Cautun}, M., {Frenk}, C.~S., {et~al.} 2016, \mnras, 460, 3772

\bibitem[{{Shi} {et~al.}(2024){Shi}, {Sunayama}, {Kurita}, {Takada},
  {Sugiyama}, {Mandelbaum}, {Miyatake}, {More}, {Nishimichi}, \&
  {Johnston}}]{Shi2024}
{Shi}, J., {Sunayama}, T., {Kurita}, T., {et~al.} 2024, \mnras, 528, 1487

\bibitem[{{Tenneti} {et~al.}(2014){Tenneti}, {Mandelbaum}, {Di Matteo}, {Feng},
  \& {Khandai}}]{Tenneti2014}
{Tenneti}, A., {Mandelbaum}, R., {Di Matteo}, T., {Feng}, Y., \& {Khandai}, N.
  2014, \mnras, 441, 470

\bibitem[{{Valenzuela} {et~al.}(2024){Valenzuela}, {Remus}, {Dolag}, \&
  {Seidel}}]{Valenzuela2024}
{Valenzuela}, L.~M., {Remus}, R.-S., {Dolag}, K., \& {Seidel}, B.~A. 2024,
  \aap, 690, A206

\bibitem[{{van Uitert} {et~al.}(2017){van Uitert}, {Hoekstra}, {Joachimi},
  {Schneider}, {Bland-Hawthorn}, {Choi}, {Erben}, {Heymans}, {Hildebrandt},
  {Hopkins}, {Klaes}, {Kuijken}, {Nakajima}, {Napolitano}, {Schrabback},
  {Valentijn}, \& {Viola}}]{vanUitert2017b}
{van Uitert}, E., {Hoekstra}, H., {Joachimi}, B., {et~al.} 2017, \mnras, 467,
  4131

\bibitem[{{van Uitert} \& {Joachimi}(2017)}]{vanUitert2017a}
{van Uitert}, E. \& {Joachimi}, B. 2017, \mnras, 468, 4502

\bibitem[{{Velliscig} {et~al.}(2015){Velliscig}, {Cacciato}, {Schaye}, {Crain},
  {Bower}, {van Daalen}, {Dalla Vecchia}, {Frenk}, {Furlong}, {McCarthy},
  {Schaller}, \& {Theuns}}]{Velliscig2015a}
{Velliscig}, M., {Cacciato}, M., {Schaye}, J., {et~al.} 2015, \mnras, 453, 721

\bibitem[{Virtanen {et~al.}(2020)Virtanen, Gommers, Oliphant, Haberland, Reddy,
  Cournapeau, Burovski, Peterson, Weckesser, Bright, {van der Walt}, Brett,
  Wilson, Millman, Mayorov, Nelson, Jones, Kern, Larson, Carey, Polat, Feng,
  Moore, {VanderPlas}, Laxalde, Perktold, Cimrman, Henriksen, Quintero, Harris,
  Archibald, Ribeiro, Pedregosa, {van Mulbregt}, \& {SciPy 1.0
  Contributors}}]{Scipy2020}
Virtanen, P., Gommers, R., Oliphant, T.~E., {et~al.} 2020, Nature Methods, 17,
  261

\bibitem[{{Wang} {et~al.}(2024){Wang}, {Mo}, {Chen}, \& {Schaye}}]{Wang2024}
{Wang}, K., {Mo}, H.~J., {Chen}, Y., \& {Schaye}, J. 2024, \mnras, 527, 10760

\bibitem[{{Wiersma} {et~al.}(2009){Wiersma}, {Schaye}, {Theuns}, {Dalla
  Vecchia}, \& {Tornatore}}]{Wiersma2009}
{Wiersma}, R. P.~C., {Schaye}, J., {Theuns}, T., {Dalla Vecchia}, C., \&
  {Tornatore}, L. 2009, \mnras, 399, 574

\bibitem[{{Zemp} {et~al.}(2011){Zemp}, {Gnedin}, {Gnedin}, \&
  {Kravtsov}}]{Zemp2011}
{Zemp}, M., {Gnedin}, O.~Y., {Gnedin}, N.~Y., \& {Kravtsov}, A.~V. 2011, \apjs,
  197, 30

\bibitem[{{Zhou} {et~al.}(2023){Zhou}, {Tong}, {Troxel}, {Blazek}, {Lin},
  {Bacon}, {Bleem}, {Chang}, {Costanzi}, {DeRose}, {Dietrich}, {Drlica-Wagner},
  {Gruen}, {Gruendl}, {Hoyle}, {Jarvis}, {MacCrann}, {Mawdsley}, {McClintock},
  {Melchior}, {Prat}, {Pujol}, {Rozo}, {Rykoff}, {Samuroff}, {Sheldon}, {Shin},
  {Rosell}, {Yanny}, {S{\'a}nchez}, {Tucker}, {Sevilla-Noarbe}, {Zuntz},
  {Varga}, {Zhang}, {Alves}, {Amon}, {Bertin}, {Brooks}, {Burke}, {Kind}, {da
  Costa}, {Davis}, {De Vicente}, {Desai}, {Diehl}, {Doel}, {Everett},
  {Ferrero}, {Flaugher}, {Frieman}, {Gerdes}, {Gutierrez}, {Hinton},
  {Hollowood}, {Honscheid}, {James}, {Jeltema}, {Kuehn}, {Lahav}, {Lima},
  {Marshall}, {Mena-Fern{\'a}ndez}, {Menanteau}, {Miquel}, {Palmese},
  {Paz-Chinch{\'o}n}, {Pieres}, {Malag{\'o}n}, {Porredon}, {Raveri}, {Romer},
  {Sanchez}, {Smith}, {Soares-Santos}, {Suchyta}, {Swanson}, {Tarle}, {To},
  {Weaverdyck}, {Weller}, \& {Wiseman}}]{Zhou2023}
{Zhou}, C., {Tong}, A., {Troxel}, M.~A., {et~al.} 2023, \mnras, 526, 323

\end{thebibliography}

\begin{appendix}
\section{Effect of mass cut on sampling bias}
\label{sect:appendixA}

\begin{figure}

    \centering
    \includegraphics[width=0.5\textwidth]{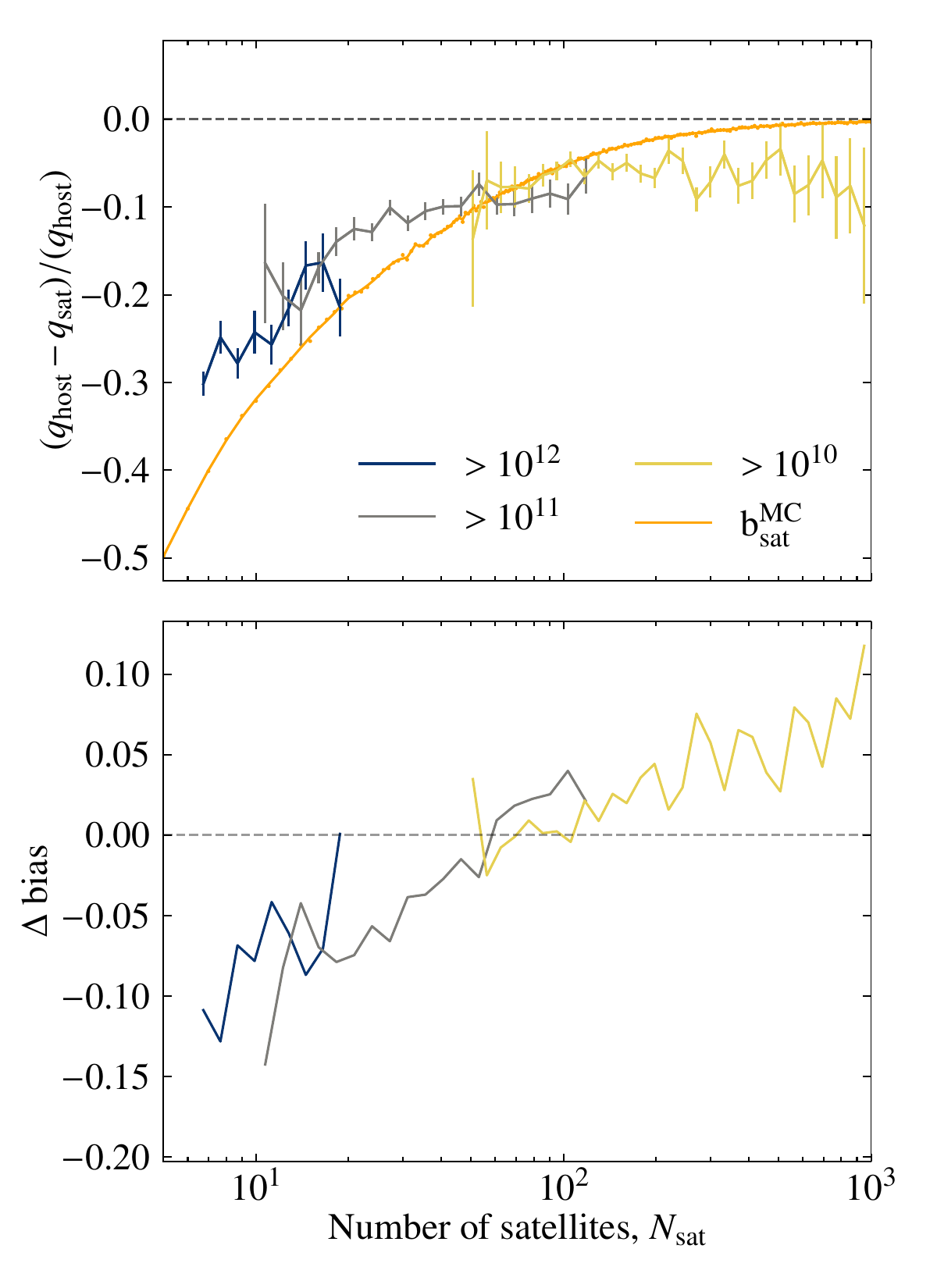}
    \caption{Top panel: Variation of projected axis ratio bias with number of satellites with different mass cuts applied to the satellite catalogue. The host halo mass is confined to haloes with DM mass $> 10^{14}$ $\mathrm{M}_{\odot}$, for the fiducial hydrodynamic $\mathrm{m}_8$ case. Bottom panel: Difference between the bias estimated from the MC and the bias estimated from the satellite positions at different mass cuts ($\Delta \ \mathrm{bias} = \mathrm{b}_{\mathrm{MC}} - \mathrm{b}_{\mathrm{sat}}$).}
    \label{fig:bias_n_sat_mass_cut}

\end{figure}

In the main body of this paper, we explored resolution effects on the shapes of haloes. In cosmological simulations, however, physical processes such as dynamical friction and relaxation also depend on resolution. These processes cause higher mass haloes to be closer to the halo centre. In addition, low survey depth in observational studies could lead to a higher effective mass cut than what we considered in this work. Thus, the effect of the mass cut we chose could potentially change the bias we measured. In this appendix, we explore the effect of different mass cuts on the estimated bias.

We first kept the sample of host haloes considered fixed, confining the analysis to those that have DM mass $> 10^{14}$ $\mathrm{M}_{\odot}$. We then estimated the axis ratio bias as we did previously, but with different satellite mass cuts of $10^{10}$, $10^{11}$ and $10^{12}$ $\mathrm{M}_{\odot}$. As expected, the number of satellites increases rapidly as the mass cut is reduced. The maximum value of $\Delta \ \mathrm{bias}$ is about 0.1 - 0.15. The value of $\Delta \ \mathrm{bias}$ does not change while increasing and decreasing the mass cut by an order of magnitude, as seen in Fig.\ref{fig:bias_n_sat_mass_cut}. We thereby conclude that the effect of the mass cut does not strongly affect our results.

\section{Orientation angle bias}
\label{sect:appendixB}

\begin{figure}
    \centering
    \includegraphics[width=0.5\textwidth]{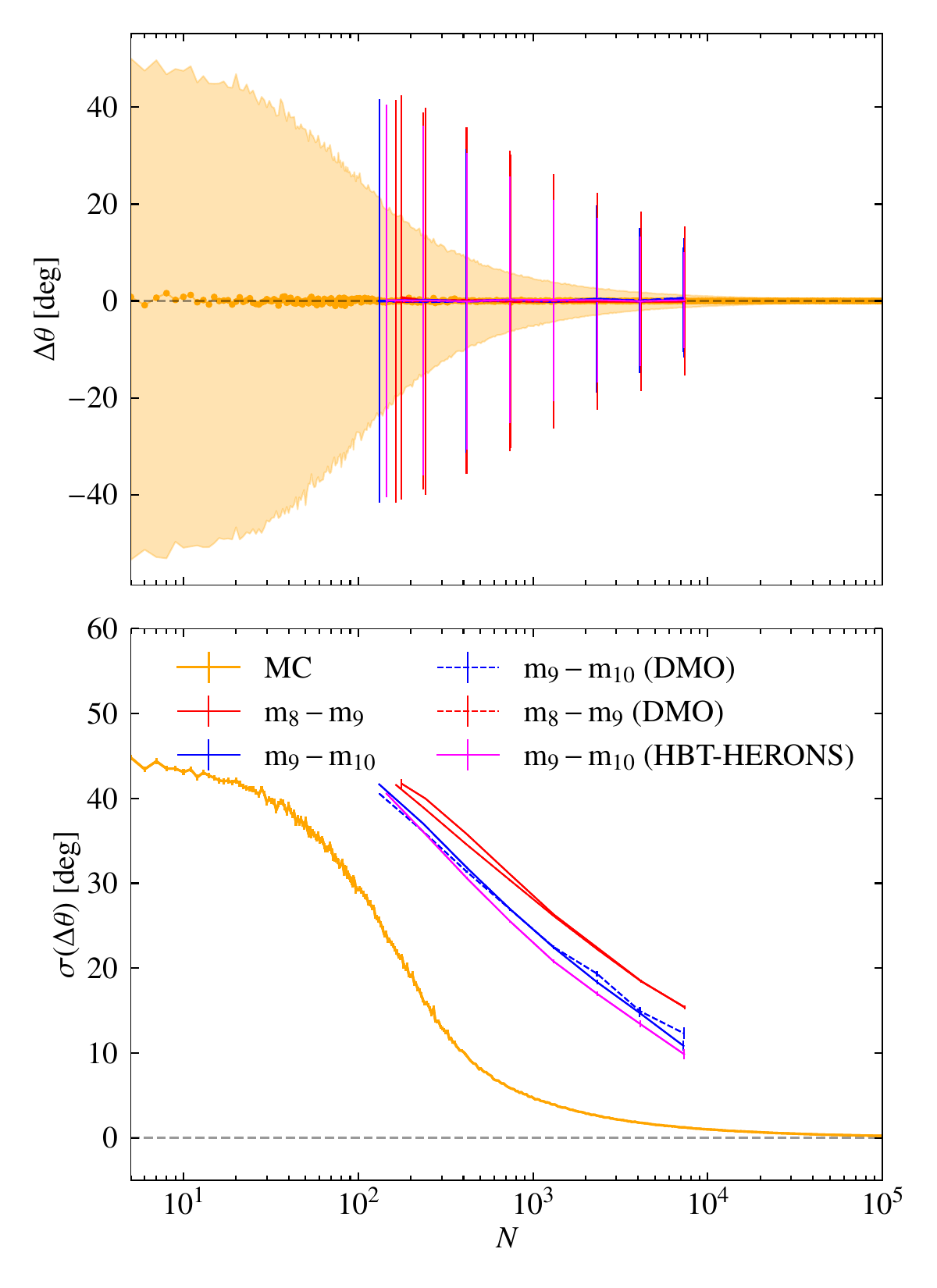}
    \caption{Top panel: The distribution of the projected misalignment angle, $\Delta \theta$, as a function of the number of particles. The orange points depict the prediction from the MC, assuming only sampling noise. The orange shaded region is the 1$\sigma$ scatter from the MC. Overplotted with error bars are the values of $\Delta \theta$ from haloes matched between the $\mathrm{m}_8$, $\mathrm{m}_9$ and $\mathrm{m}_{10}$ resolution runs of the simulation. Bottom panel: The scatter of $\Delta \theta$ as a function of the number of particles. This is the scatter of the top panel, with the r.m.s errors calculated using bootstrapping. The scatter on the misalignment angle from the simulations are larger than our pure sampling noise model, due to the additional scatter arising when resolved low-mass substructure is excised from the host halo.}
    \label{fig:position_angle_bias}
\end{figure}

The main theme of this paper has been to explore the effect of resolution on the derived axis ratios of haloes and we showed that the bias arising from sampling noise can be modeled using our MC. A logical extension is the application of this approach to the orientation angles of haloes. In this appendix, we will explore this further and the challenges associated with it.

\begin{figure*}
  \centering
  \subfloat[a][]{\includegraphics[width=0.33\textwidth]{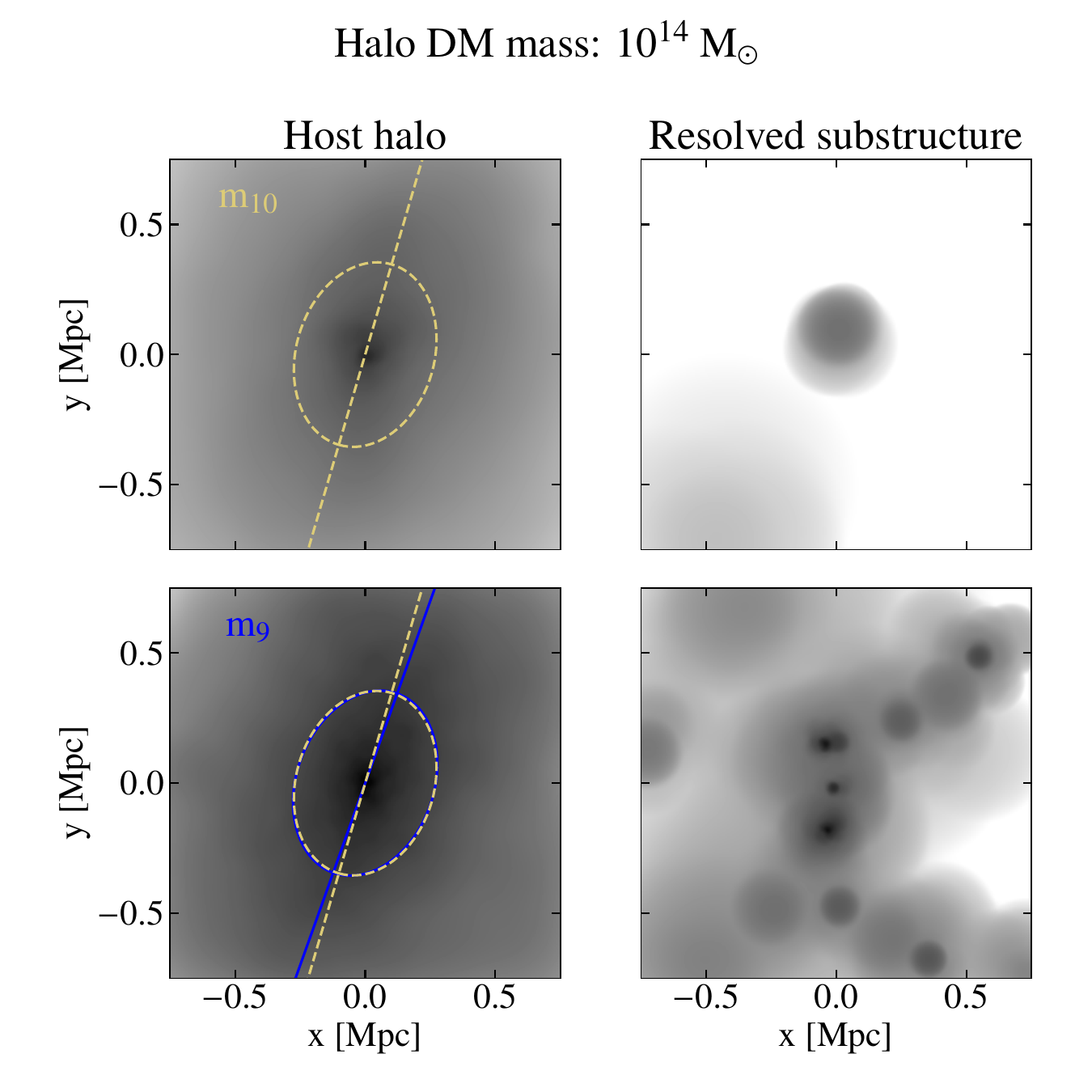} \label{fig:misaligned_example1}}
  \subfloat[b][]{\includegraphics[width=0.33\textwidth]{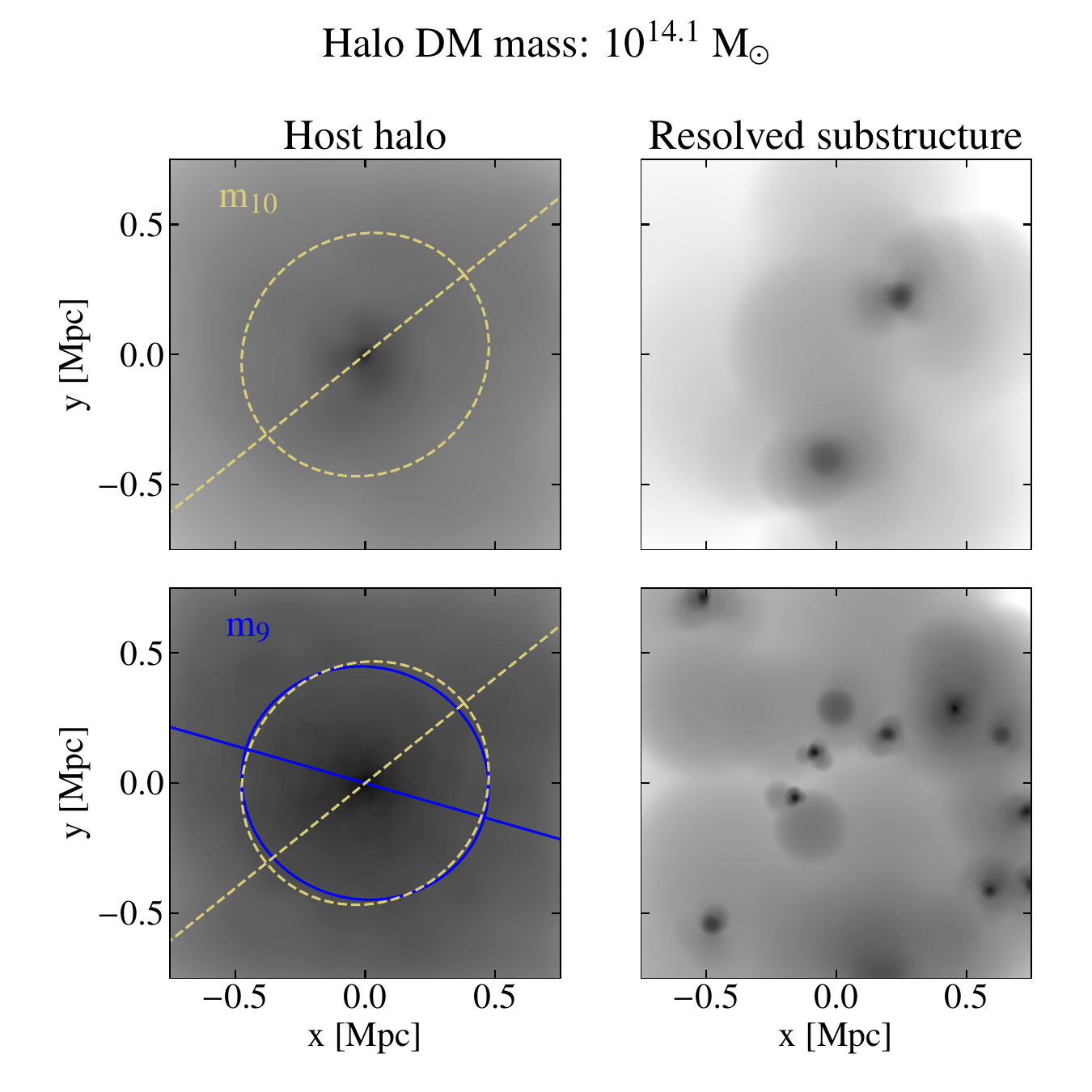} \label{fig:misaligned_example2}}
  \subfloat[c][]{\includegraphics[width=0.33\textwidth]{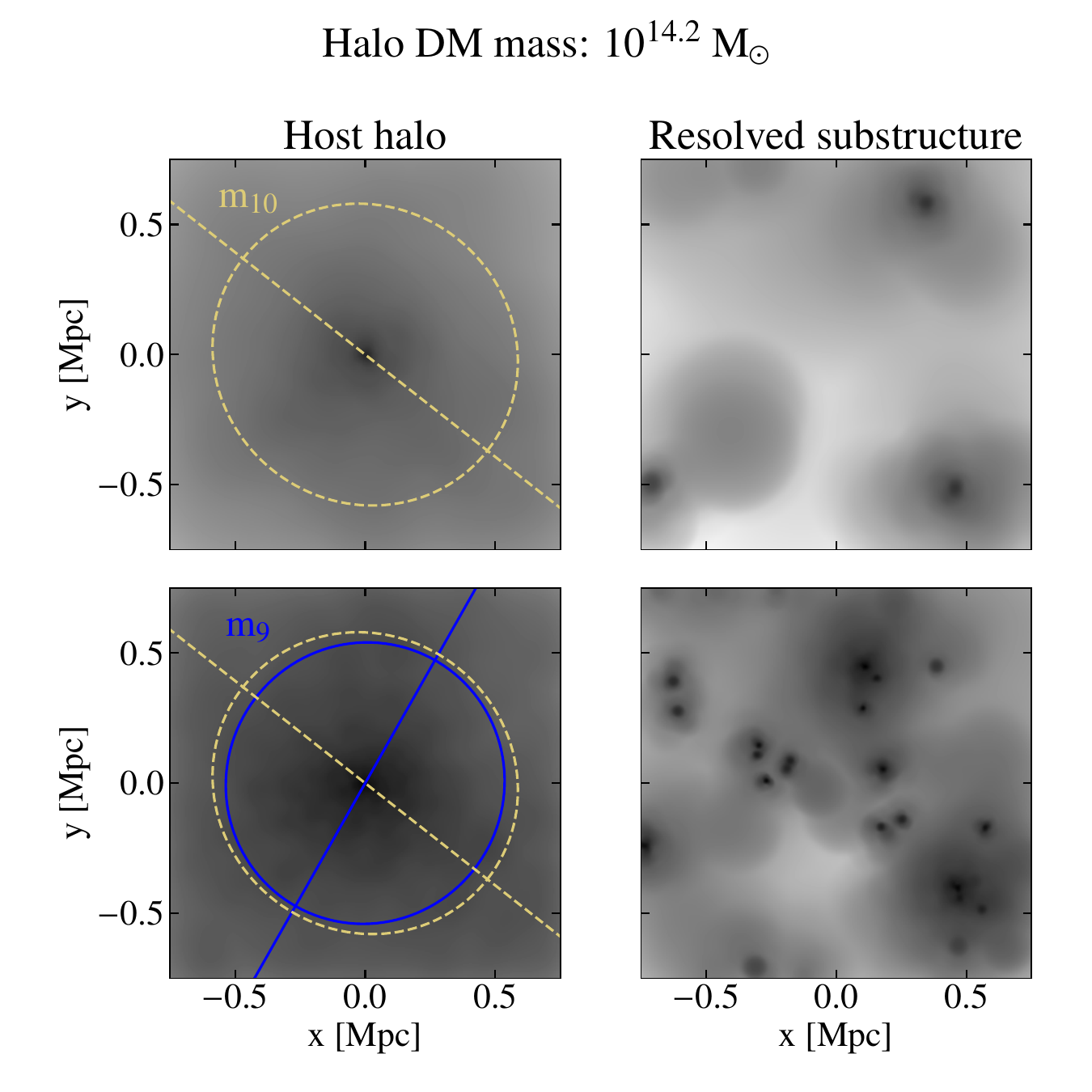} \label{fig:misaligned_example3}}
  \caption{Examples of matched haloes in the $\mathrm{m}_{10}$ run (upper row) and the $\mathrm{m}_{9}$ run (bottom row). The left panels of figures (a), (b) and (c) show the host halo with the projected ellipse and orientation angle overplotted, with the right panels showing the resolved substructures that are excised from the host halo before calculating the inertia tensor. This sources an additional scatter over that from sampling noise. (a) shows an example for which the effect is relatively small, causing a misalignment of only a few degrees. (b) is an example of a more spherical halo that is hence more sensitive to the removal of low mass substructures. (c) is another spherical halo for which the orientation angle of the $\mathrm{m}_{9}$ run is almost orthogonal to that from the $\mathrm{m}_{10}$ run.}
  \label{fig:misaligned_example}
\end{figure*}

First, we define $\Delta \theta$, the projected misalignment in orientation angles between two haloes that have different numbers of particles. By definition, $\Delta \theta$ is centered around 0 and has a scatter, $\sigma(\Delta \theta)$ associated with it. This scatter captures how much misalignment arises purely from resolution effects, i.e., having too few particles to sample the orientation angle of the ellipse. $\Delta \theta$ is plotted in the top panel of Fig.\ref{fig:position_angle_bias} with the associated scatter shown in the bottom panel. For the MC, the scatter is large for low $N$, and quickly approaches 0 as $N$ is increased. When we compare against matched haloes in the different resolution runs, however, we see that the values calculated from the simulation are consistently higher at all $N$. 

This can be explained by the fact that, in addition to the scatter from sampling noise, there is a resolution effect from the act of halo finding that is not captured in our MC. This can be understood as follows: when a group of particles have a sufficient density contrast compared to the main halo, they are considered a separate object. Once these substructures are found, they are excised from the dark matter distribution before calculating the inertia tensors of the host. Increasing the particle resolution means more substructure is associated with each host, and thus comparing between two resolution runs captures this additional scatter as well, which is not represented in the MC. Even at $10^4$ particles, there is a scatter of around 10$^{\degree}$.

We show this effect qualitatively in Fig.\ref{fig:misaligned_example}. The top panel of each figure shows the dark matter distribution of a halo in the $\mathrm{m}_{10}$ run within a $1.5 \times1.5 \ \mathrm{Mpc}$ region centered on the halo centre of potential, and the bottom panel shows the matched halo in the $\mathrm{m}_{9}$ run. The left panels show the host dark matter halo with the projected ellipse and orientation angle overplotted. The right plots show objects that are not bound to the host halo, but are distinct substructures (`Resolved substructure'). In each example, there are more substructures in the $\mathrm{m}_{9}$ run, which are excised from the host dark matter distribution before the inertia tensors are calculated. 

For the example halo with dark matter mass of $10^{14} \ \mathrm{M}_{\odot}$ in Fig.\ref{fig:misaligned_example1} (a), the scatter due to both the sampling noise and the removal of low-mass substructure is small. For some haloes, however, the effect is more extreme, causing misalignments of more than $45^{\degree}$, shown in Figs. \ref{fig:misaligned_example} (b) and (c). This is because these haloes are very spherical and hence very sensitive to the effect of substructure removal.

The effect of removing low-mass structures is not captured in our MC, which results in the mismatch when compared to the multiple resolution runs from the FLAMINGO simulations. This can affect the measurements of the intrinsic alignment projected correlation function $w_{\mathrm{g}+}$ at small scales, as pointed out by \citet{Chisari2016} (see their Section 4.2.2). The additional scatter from the removal of low-mass subhaloes cannot be captured by the traditional method of downsampling either, as halo finding is not carried out at each iteration. In order to correctly account for this effect, one would need to make mock NFW haloes as was done in our MC, with the addition of substructure following a realistic number density at each mass as well as spatial distributions. Distinct subhaloes would need to be found at each value of $N$, which would capture this additional scatter. This is beyond the scope of this work.

\end{appendix}

\end{document}